\documentclass[twocolumn,aps,showpacs,superscriptaddress]{revtex4-1}

\usepackage{graphicx}

\begin{document}

%\title{Higher-order non-classicality criteria for optical fields based on moments from quadratic detectors}
\title{Higher-order sub-Poissonian-like nonclassical fields: Theoretical and
experimental comparison}

\author{Jan Pe\v{r}ina Jr.}
\email{jan.perina.jr@upol.cz} \affiliation{RCPTM, Joint Laboratory
of Optics of Palack\'{y} University and Institute of Physics of
the Czech Academy of Sciences, Faculty of Science, Palack\'{y}
University, 17. listopadu 12, 77146 Olomouc, Czech Republic}

\author{V\'{a}clav Mich\'{a}lek}
\affiliation{Institute of Physics of the Czech Academy of
Sciences, Joint Laboratory of Optics of Palack\'{y} University and
Institute of Physics of CAS, 17. listopadu 50a, 772 07 Olomouc,
Czech Republic}

\author{Ond\v{r}ej Haderka}
\affiliation{RCPTM, Joint Laboratory of Optics of Palack\'{y}
University and Institute of Physics of the Czech Academy of
Sciences, Faculty of Science, Palack\'{y} University, 17.
listopadu 12, 77146 Olomouc, Czech Republic}

\begin{abstract}
Criteria defining higher-order sub-Poissonian-like fields are
given using five different quantities: moments of I) integrated
intensity, II) photon number, III) integrated-intensity
fluctuation, IV) photon-number fluctuation, and V) elements of
photocount and photon-number distributions. Relations among the
moment criteria are revealed. Performance of the criteria is
experimentally investigated using a set of potentially
sub-Poissonian fields obtained by post-selection from a twin beam.
The criteria based on moments of integrated intensity and photon
number and those using the elements of photocount distribution are
found as the most powerful. States nonclassical up to the fifth
order are experimentally reached in the former case, even the
ninth-order non-classicality is observed in the latter case.
\end{abstract}

% \pacs{42.65.Lm,42.50.Ar}
% 42.65.Lm    Parametric down conversion and production of entangled photons
% 42.50.Ar    Photon statistics and coherence theory

\maketitle

Nonclassical properties of optical fields and their
characterization have been in the center of attention from the
beginning of quantum optics. The simplest, and from the
experimental point of view the most natural, way how to achieve
this is based on the determination of second-order moments of
fluctuations of the measured quantities, that violate certain
inequalities for nonclassical fields. This approach resulted in
the introduction of principal squeeze variance of electric-field
amplitudes and the Fano factor to quantify nonclassical phase
fluctuations and photon-number fluctuations, respectively
\cite{Luks1988,Davidovich1996}. The Fano factor represents the
most important quantity for optical fields characterized by
standard quadratic detectors, for which it identifies
sub-Poissonian fields. It has been used to quantify nonclassical
light originating in resonance fluorescence
\cite{Kimble1977,Short1983}, Franck--Hertz experiment
\cite{Teich1985}, high-efficiency light-emitting diodes
\cite{Tapster1987}, second-harmonic generation
\cite{Li1994,Bajer1999}, parametric deamplification \cite{Li1995},
second-subharmonic generation \cite{Koashi1993}, feed-forward
action on the beam \cite{Mertz1990,Kim1992} or light generated in
micro-cavities by passing atoms \cite{Raimond2001}. Highly
sub-Poissonian fields have also been reached by post-selection
from cw \cite{Rarity1987,Laurat2003,Zou2006} and pulsed twin beams
(TWB)
\cite{Bondani2007,PerinaJr2013b,Lamperti2014,Iskhakov2016,Harder2016}.

The Fano factor $ F $ defined in terms of photon-number moments as
$ F = \langle (\Delta \hat{n})^2\rangle /\langle \hat{n} \rangle $
identifies sub-Poissonian fields if $ F<1 $; $ \Delta \hat{n}
\equiv \hat{n} - \langle \hat{n}\rangle $ denotes the fluctuation
of photon-number operator $ \hat n $ given in terms of the
annihilation ($ \hat a $) and creation ($ \hat{a}^\dagger $)
operators as $ \hat n \equiv \hat{a}^\dagger \hat{a} $. Symbol $
\langle\rangle $ stands for the mean value. This condition when
expressed in the moments of integrated intensity $ W $ (or
equivalently in the normally-ordered moments of photon number,
i.e. $ \langle W^k\rangle \equiv \langle \hat{a}^{\dagger k}
\hat{a}^k \rangle $ \cite{Perina1991,Saleh1978,Vogel2006}), $
\langle (\Delta W)^2 \rangle = \langle \hat{a}^{\dagger 2}
\hat{a}^2 \rangle - \langle \hat{a}^{\dagger} \hat{a} \rangle^2 <
0 $ [for the relation between the moments that is used for
determining intensity moments from the experimental data, see
Eq.~(\ref{3}) below], reveals the relation with the general
definition of non-classicality: A field is nonclassical provided
that its (normally-ordered) Glauber-Sudarshan quasi-distribution $
{\cal P} $ (as a function of complex field amplitudes) attains
negative values or even does not exist as a regular function
\cite{Glauber1963,Sudarshan1963}. The consideration of the
marginal quasi-distribution $ P $ of integrated intensities,
application of this definition to any classical field and use of
the Cauchy-Schwarz inequality (or the majorization theory
\cite{Lee1990a}) result in the chain of inequalities $ \langle
W^k\rangle > \langle W\rangle^k $ fulfilled by any classical
field. These inequalities then allow to naturally define a $ k
$-th order non-classicality (with respect to intensity $ W $)
\cite{Mista1977,Perina1991,Hong1985b,Prakash2006,Verma2010}
according to the following \emph{Criteria I}:
\begin{equation}   % 1
 r_{W}^{(k)} \equiv \langle W^k\rangle/\langle W\rangle^k  - 1
  < 0, \hspace{5mm} k=2,\ldots .
\label{1}
\end{equation}
As the quasi-distribution $ P $ of integrated intensity completely
describes the field intensity, we consider the definition
(\ref{1}) of higher-order non-classicalities as the most
fundamental. We note that different kinds of higher-order
non-classicalities have been defined when considering powers of
complex field amplitudes
\cite{Hong1985a,Hillery1987a,Hillery1987b}.

On the other hand, photon-number-resolving detectors
straightforwardly provide the moments of photon number $ \hat{n}
$. The following sequence of non-classicality \emph{Criteria II}
can be defined using these moments:
\begin{equation}  % 2
 r_{n}^{(k)} \equiv \langle \hat{n}^k\rangle / \langle \hat{n}^k\rangle_{\rm Pois} - 1
   < 0, \hspace{5mm} k=2,\ldots.
\label{2}
\end{equation}
The moments $ \langle \hat{n}^k\rangle_{\rm Pois} $ characterize a
Poissonian field (in a coherent state) with mean photon number $
\langle \hat{n}\rangle $. Indeed, the relation among both types of
moments expressed via the Stirling numbers $ S_k^l $ of the second
kind ($ k \ge 1 $),
\begin{equation}  % 3
 \langle \hat{n}^k \rangle = \sum_{l=1}^{k} S_k^l \langle W^l \rangle, \hspace{3mm}
  S_k^l \equiv \frac{1}{l!} \sum_{m=0}^{l} (-1)^{l-m}
  \left( \begin{array}{c} l \\ m \end{array} \right) m^k,
%  \hspace{0mm} k=1,\ldots,
\label{3}
\end{equation}
allows to rewrite criteria (\ref{2}) into the form:
\begin{equation}  % 4
 \langle \hat{n}^k\rangle - \langle \hat{n}^k\rangle_{\rm Pois} =
  \sum_{l=1}^{k} S_k^l \left( \langle W^k\rangle - \langle
  W\rangle^k \right) < 0.
\label{4}
\end{equation}

The relation (\ref{4}) together with positivity of the Stirling
numbers $ S $ confirm that criteria (\ref{2}) express
non-classicality. Whereas Criteria I in Eqs.~(\ref{1}) and II in
Eqs.~(\ref{2}) are identical for $ k=2 $, they represent in
general different definitions of a $ k $-th order
non-classicality. For example, a field obeying $ \langle W^3
\rangle - \langle W\rangle^3 < 0 $ does not have to fulfill the
condition $ \langle \hat{n}^3 \rangle - \langle
\hat{n}^3\rangle_{\rm Poiss} < 0 $ and vice versa. Both Criteria I
and II approach each other only for intense fields ($ \langle
W\rangle \gg 1 $) for which the last term in the sum in
Eq.~(\ref{4}) dominates ($ S_k^k = 1 $ for $ k=1,\ldots $).

Non-classicality of an optical field can also be revealed by the
moments of intensity ($ \Delta W \equiv W - \langle W \rangle $)
and photon-number ($ \Delta \hat{n}  \equiv \hat{n} - \langle
\hat{n} \rangle $) fluctuations. This leads us to the following
\emph{Criteria III} and \emph{IV}:
\begin{eqnarray}   % 5,6
 r_{\Delta W}^{(k)} &\equiv& \langle (\Delta W)^k\rangle / \langle
   W\rangle^k,
  \label{5} \\
 r_{\Delta n}^{(k)} &\equiv& \langle (\Delta\hat{n})^k\rangle / \langle (\Delta \hat{n})^k\rangle_{\rm Pois} - 1,
  \hspace{5mm} k=2,\ldots .
\label{6}
\end{eqnarray}
We note that $ \langle (\Delta \hat{n})^2 \rangle_{\rm Pois} =
\langle (\Delta \hat{n})^3 \rangle_{\rm Pois} = \langle \hat{n}
\rangle $, $ \langle (\Delta \hat{n})^4 \rangle_{\rm Pois} =
\langle \hat{n} \rangle + 3\langle \hat{n} \rangle^2 $ and $
\langle (\Delta \hat{n})^5 \rangle_{\rm Pois} = \langle \hat{n}
\rangle + 10\langle \hat{n} \rangle^2 $. However, Criteria III $
r_{\Delta W}^{(k)} < 0 $ for intensity fluctuations are applicable
only for even orders $ k $. Also Criteria IV $ r_{\Delta n}^{(k)}
< 0 $ reveal non-classicality only for fields with mean
intensities $ \langle W \rangle $ lower than certain value. A
detailed analysis of expressions $ \langle
(\Delta\hat{n})^k\rangle - \langle (\Delta \hat{n})^k\rangle_{\rm
Pois} $ rewritten as polynomials of $ k $-th order in $ \Delta W $
with $ \langle W \rangle $ considered as a parameter
\cite{Kim1998} gives $ r_{\Delta n}^{(3)} < 0 $ as a
non-classicality indicator for $ \langle W \rangle < 3 $ and $
r_{\Delta n}^{(4)} < 0 $ for arbitrary intensities. % $ \langle W \rangle $.

Formally similar non-classicality criteria are derived for the
elements $ p(k) $ of photon-number distribution
\cite{Klyshko1996}. These elements, given by the Mandel detection
formula \cite{Perina1991,Saleh1978}
\begin{equation}  % 7
 p(k) = \int_{0}^{\infty} dW \, W^k \exp(-W) P(W) /k!,
\label{7}
\end{equation}
represent 'un-normalized' moments that obey, according to the
majorization theory, certain inequalities for non-negative
distribution $ P(W) $ of integrated intensity
\cite{Klyshko1996,Lee1998}. Nonclassical fields are then
identified by their violation, which allows us to formulate
\emph{Criteria V} in terms of the modified elements $ \tilde{p}(k)
\equiv k! p(k)/p(0) $, $ k=1,2,\ldots $, as follows:
\begin{equation}  % 8
 r_p^{(k)} \equiv \tilde{p}(k) / \tilde{p}(1)^k - 1 < 0,
  \hspace{5mm} k=2,\ldots .
\label{8}
\end{equation}
For a Poissonian field, we have $ r_p^{(k)} = 0 $ for all $ k $.
Moreover, the majorization theory allows to derive a larger number
of non-classicality inequalities among the elements $ \tilde{p}(k)
$, in tight parallel with those for intensity moments $ \langle
W^k \rangle $ analyzed for $ k \le 5 $ in \cite{Arkhipov2016c}.

\emph{Criterion I} $ r_{W}^{(k)} < 0 $ for a $ k $-th order
non-classicality can be converted into the $ k $-th-order
non-classicality depth $ \tau^{(k)} $ \cite{Lee1991} using the
formula
\begin{equation}    % 9
 \tau^{(k)} = (1-s^{(k)}_{\rm th})/2 ,
\label{9}
\end{equation}
where $ s^{(k)}_{\rm th} $ gives the threshold value of the
ordering parameter $ s $ for which $ \langle W^k \rangle_s =
\langle W\rangle^k_s $. The $ s $-ordered moments $ \langle W^k
\rangle_s $ of intensity $ W $, which are determined along the
usual way considering an $ s $-ordered quasi-distribution $
\tilde{P}(W;s) $ of integrated intensity [see below], are
expressed in terms of the usual normally-ordered moments ($ s=1 $)
as follows \cite{Perina1991,Vogel2006}:
\begin{equation}   % 10
 \langle W^k\rangle_s = \left(\frac{2}{1-s}\right)^k \left\langle
  {\rm L}_k \left(\frac{2W}{s-1}\right) \right\rangle ;
\label{10}
\end{equation}
$ {\rm L}_k $ denotes a $ k $-th Laguerre polynomial
\cite{Gradshtein2000}. We note that such moments are appropriate
for a field into which a thermal field with $ (1-s) $ mean photon
number is added. Contrary to the parameters $ r_{W}^{(k)} $ the
non-classicality depths $ \tau^{(k)} $ of different orders can be
mutually directly compared. The greater the value of $ \tau^{(k)}
$ is the stronger the non-classicality is.

The parameters $ r_{W}^{(k)} $ naturally occur when determining
the declination $ \Delta \tilde P(W;s) $ of an $ s $-ordered
quasi-distribution $ \tilde P(W;s) $ of integrated intensity from
that belonging to the Poissonian field, which is denoted as $
\tilde P_{\rm Pois}(W;s) $ \cite{Perina1991}:
\begin{eqnarray}   % 11
 \Delta \tilde P(W;s)&=& \exp\left(-\frac{W}{\langle W\rangle}\right)
  \sum_{j=0}^{\infty} c_j {\rm L}_j \left(-\frac{W}{\langle
  W\rangle}\right), \nonumber \\
  c_j &=& \frac{j!}{\langle W\rangle} \sum_{l=0}^{j} \frac{ (-1)^l
   r_{W}^{(l)} }{(l!)^2 (j-l)! } .
\label{11}
\end{eqnarray}
It holds that $ \int_{0}^{\infty} dW \Delta \tilde P(W;s) = 0 $
and so $ \Delta \tilde P(W;s) $ of any non-Poissonian field has to
have negative values. However, negative values of a non-classical
field occur in the regions where they cannot be compensated by
positive values of the Poissonian distribution $ \tilde P_{\rm
Pois}(W;s) $.

The performance of different non-classicality quantifiers has been
experimentally tested on a set of 10 potentially sub-Poissonian
fields obtained by post-selection from a TWB. The used TWB was
generated in a nonlinear crystal and its signal and idler fields
were detected in different regions of an iCCD camera
\cite{PerinaJr2012} (for details, see Fig.~1). The signal
photocounts were used for the post-selection process: Detection of
a given number $ c_{\rm s} $ of signal photocounts ideally leaves
the idler field in the state with $ c_{\rm i}=c_{\rm s} $ idler
photons. Under real experimental conditions, the post-selected
idler field exhibits fluctuations in photon numbers that, however,
are under suitable conditions smaller than those characterizing
the corresponding Poissonian field. The post-selected idler fields
were measured via their photocount distributions monitored by the
iCCD camera. The obtained post-selected idler fields had different
intensities as the mean number $ \langle n_{\rm i} \rangle $ of
idler photons increases with the increasing signal photocount
number $ c_{\rm s} $.

Moreover, as the experiment provided the whole 2D joint
signal-idler photocount histogram $ f(c_{\rm s},c_{\rm i}) $ it
also allowed to reconstruct the whole TWB. The TWB was
reconstructed as a field composed of three independent components,
one characterizing ideal photon pairs, one describing noisy signal
photons and one belonging to noisy idler photons. Each component
is characterized by mean photon(-pair) number $ B_a $ per mode and
number $ M_a $ of independent modes, $ a={\rm p,s,i} $, and its
photon-number distribution is given by the Mandel-Rice formula
\cite{Perina1991,Saleh1978,PerinaJr2013a}. The distribution $
p_{\rm si}(n_{\rm s},n_{\rm i}) $ of the whole TWB is then
expressed in the form of the following two-fold convolution
\cite{PerinaJr2012a,PerinaJr2013a,Perina2005}:
\begin{eqnarray}  % 12
 p_{\rm si}(n_{\rm s},n_{\rm i}) &=& \sum_{n=0}^{{\rm min}[n_{\rm s},n_{\rm i}]}
  p(n_{\rm s}-n;M_{\rm s},B_{\rm s})
  p(n_{\rm i}-n;M_{\rm i},B_{\rm i}) \nonumber \\
 & & \mbox{} \times  p(n;M_{\rm p},B_{\rm p});
\label{12}
\end{eqnarray}
$ p(n;M,B) = \Gamma(n+M) / [n!\, \Gamma(M)] B^n/(1+B)^{n+M} $ and
symbol $ \Gamma $ denotes the $ \Gamma $-function.

For the reconstructed TWB, the theoretical post-selected idler
photon-number distributions $ p_{\rm c,i}^{\rm theo}(n_{\rm
i};c_{\rm s}) $ observed after detecting $ c_{\rm s} $ signal
photocounts are expected in the form (for details, see
\cite{PerinaJr2013b}):
\begin{equation}   % 13
 p_{\rm c,i}^{\rm theo}(n_{\rm i};c_{\rm s}) = \frac{
  \sum_{n_{\rm s}} T_{\rm s}(c_{\rm s},n_{\rm s}) p_{\rm si}(n_{\rm s},n_{\rm i}) }{
  f_{\rm s}^{\rm theo}(c_{\rm s}) }
\label{13}
\end{equation}
where $ f_{\rm s}^{\rm theo}(c_{\rm s}) \equiv \sum_{n_{\rm s},
n_{\rm i}} T_{\rm s}(c_{\rm s},n_{\rm s}) p_{\rm si}(n_{\rm
s},n_{\rm i}) $ is the expected signal-field photocount
distribution. Function $ T_{\rm s}(c_{\rm s},n_{\rm s}) $
occurring in Eq.~(\ref{13}) characterizes detection by the camera:
It determines the probabilities of having $ c_{\rm s} $
photocounts when detecting a field with $ n_{\rm s} $ photons. For
the used iCCD camera and both detection areas with $ N_a $ active
pixels, detection efficiencies $ \eta_a $ and mean dark counts per
pixel $ D_a $, $ a={\rm s,i} $, we have \cite{PerinaJr2012}:
\begin{eqnarray}     % 14
  T_a(c_a,n_a) &=& \left(\begin{array}{c} N_a \\ c_a \end{array}\right) (1-D_a)^{N_a}
   (1-\eta_a)^{n_a} (-1)^{c_a} \nonumber \\
  & &  \mbox{} \hspace{-15mm} \times  \sum_{l=0}^{c_a} \left(\begin{array}{c} c_a \\ l \end{array}\right)
    \frac{(-1)^l}{(1-D_a)^l}  \left( 1 + \frac{l}{N_a} \frac{\eta_a}{1-\eta_a}
   \right)^{n_a}. %; \hspace{3mm} a={\rm s,i}.
\label{14}
\end{eqnarray}

In the experiment, the photon-number distributions of the
post-selected idler fields were reached by applying the
maximum-likelihood approach (MLA) \cite{Dempster1977}. The
photon-number distribution $ p_{\rm c,i}(n_{\rm i};c_{\rm s}) $
conditioned by detection of $ c_{\rm s} $ signal photocounts has
been found as a steady state of the following iteration procedure
\cite{PerinaJr2012}
\begin{eqnarray} % 15
 p_{\rm c,i}^{(l+1)}(n_{\rm i};c_{\rm s}) &=& p_{\rm c,i}^{(l)}(n_{\rm i};c_{\rm s})
  \sum_{c_{\rm i}} \frac{ f_{\rm i}(c_{\rm i};c_{\rm s})
  T_{\rm i}(c_{\rm i},n_{\rm i}) }{ \sum_{n'_i} T_{\rm i}(c_{\rm i},n'_{\rm i})
  p_{\rm c,i}^{(l)}(n'_{\rm i};c_{\rm s}) }, \nonumber \\
 & & \hspace{15mm} l=0,1,\ldots .
\label{15}
\end{eqnarray}
In Eq.~(\ref{15}), the normalized idler-field 1D photocount
histograms $ f_{\rm i}(c_{\rm i};c_{\rm s}) \equiv f(c_{\rm
s},c_{\rm i})/f_{\rm s}(c_{\rm s}) $ with the signal photocount
histogram $ f_{\rm s}(c_{\rm s}) \equiv \sum_{c_{\rm i}}f(c_{\rm
s},c_{\rm i}) $ include the experimental realizations with $
c_{\rm s} $ observed signal photocounts.
\begin{figure}  % fig. 1
 \centerline{\resizebox{0.8\hsize}{!}{\includegraphics{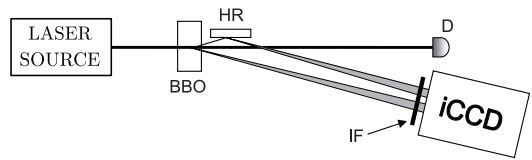}}}
 \caption{Scheme of the experimental setup: A TWB was emitted in non-collinear geometry in a
 5-mm-long type-I BaB$ {}_2 $O$ {}_4 $ crystal (BBO) pumped by the third
 harmonics (280~nm) of a femtosecond cavity dumped Ti:sapphire laser (pulse
 duration 150~fs, central wavelength 840~nm, rep. rate 50 kHz, power 5~mW, collimated 1.5-mm-wide beam).
 The signal and idler (after reflection on a highly-reflecting mirror HR)
 fields generated by a single pump pulse were detected with detection efficiencies $ \eta_{\rm s}=0.230\pm 0.005 $ and
 $ \eta_{\rm i}=0.220\pm 0.005  $
 by $N_{\rm s}=6528$ and $N_{\rm i}=6784$ pixels of the photocathode
 of iCCD camera Andor DH334-18U-63 with mean dark counts per pulse $ d = 0.04 $ ($ D_a=d/N_a$, $ a={\rm s,i}
 $), rep. rate 10 Hz and integration time 4~ns.
 The nearly-frequency-degenerate signal and
 idler photons at the wavelength of 560~nm were filtered by
 a 14-nm-wide bandpass interference filter IF that defined the
 measured TWB with parameters $M_{\rm p}=270$,
 $B_{\rm p}=0.032$, $M_{\rm s}=0.01$, $B_{\rm s}=7.6$, $M_{\rm i} = 0.026$,
 and $B_{\rm i}=5.3 $ determined with relative error 7~\% (for
 details, see \cite{PerinaJr2013a}); $\langle c_{\rm s}\rangle=2.20 \pm 0.01 $ and $\langle c_{\rm
 i}\rangle=2.18\pm 0.01$. Pump-beam intensity was actively
 stabilized via a motorized half-wave plate followed by a polarizer and monitored by
 detector D during $1.2\times 10^6$ repetitions of the measurement.}
\end{figure}

The experimental post-selection procedure provided ten different
idler fields with different probabilities $ f_{\rm s}(c_{\rm s}) $
of realization [see Fig.~2(a)] and mean idler photon numbers $
\langle n_{\rm c,i} \rangle $ increasing from 7 to 15 as the
signal photocount number $ c_{\rm s} $ used in the post-selection
increases [see Fig.~2(b)]. In Fig.~2, and also in subsequent
figures, we plot the quantities related to experimental
photocounts (photon numbers reached by MLA) by red asterisks
(green triangles) and those originating in the Gaussian fit of the
TWB (GTWB) by blue solid curves. For comparison, we also depict by
brown diamonds quantities reached by the simplest reconstruction
method based on the intensity moments and relations $ \langle W^k
\rangle \rightarrow \langle W^k \rangle / \eta^k $. As this method
does not take into account dark counts, it overestimates in
general photon-number moments of the reconstructed distributions,
as illustrated for the mean idler photon numbers $ \langle n_{\rm
i} \rangle $ in Fig.~2(b). As the experimental errors are linearly
proportional to $ 1/\sqrt{N_{\rm rep}} $ with $ N_{\rm rep} $
giving the number of measurement repetitions and according to the
graph in Fig.~2(a), the characterization of the post-selected
idler photon-number distributions with the signal photocount
numbers $ c_{\rm s} $ greater than 7 suffers from larger errors
due to the low numbers of appropriate measurements, despite the
large number of $ 1.2 \times 10^6 $ overall measurements made. We
note that fixed detection efficiencies $ \eta_{\rm s} $ and $
\eta_{\rm i} $ were considered when determining the experimental
errors of the analyzed criteria as their possible variations have
only negligible influence to identification of non-classicality in
the reconstructed states.
\begin{figure}  % fig. 2
 \centerline{\resizebox{0.5\hsize}{!}{\includegraphics{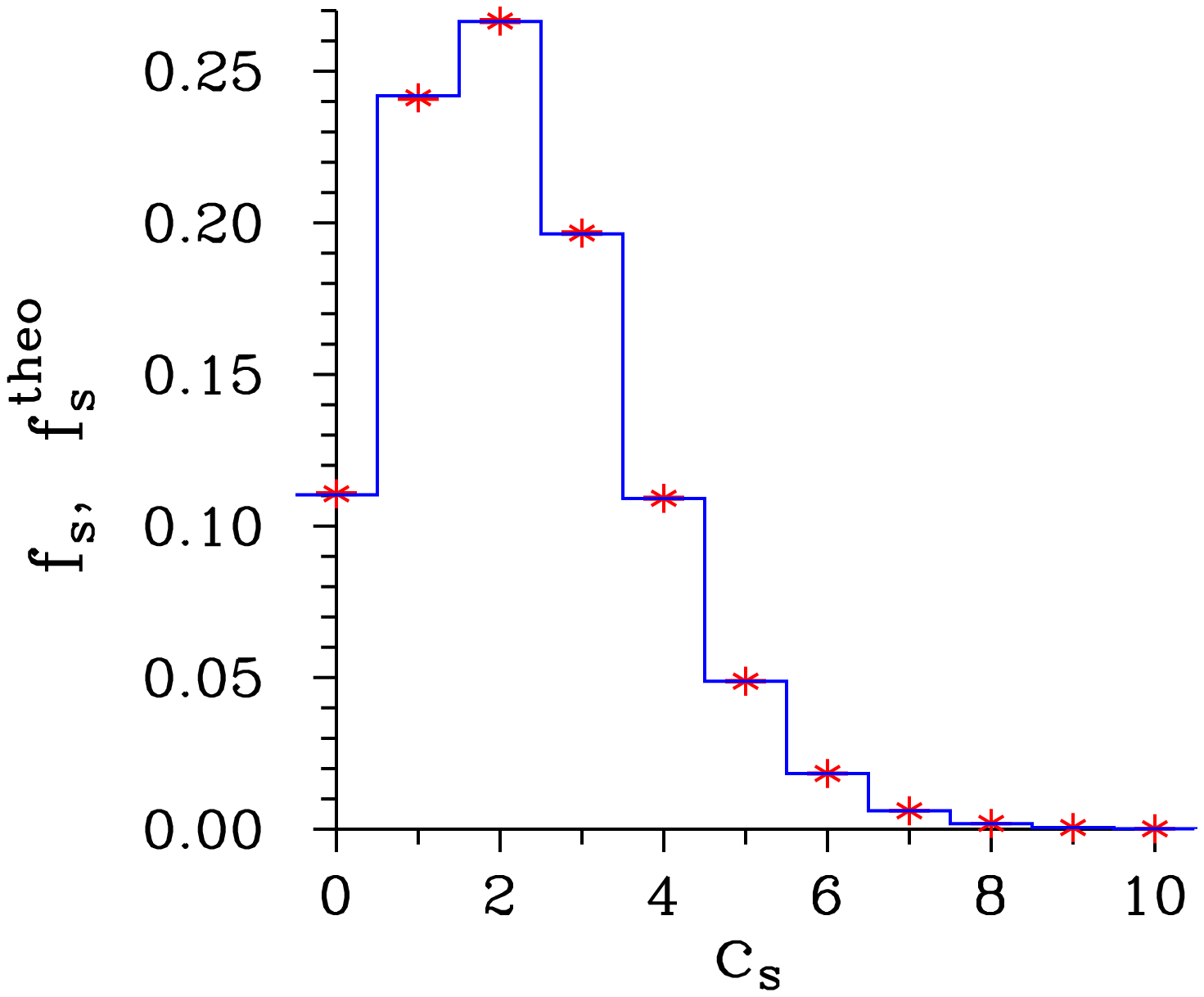}}
  \hspace{1mm} \resizebox{0.46\hsize}{!}{\includegraphics{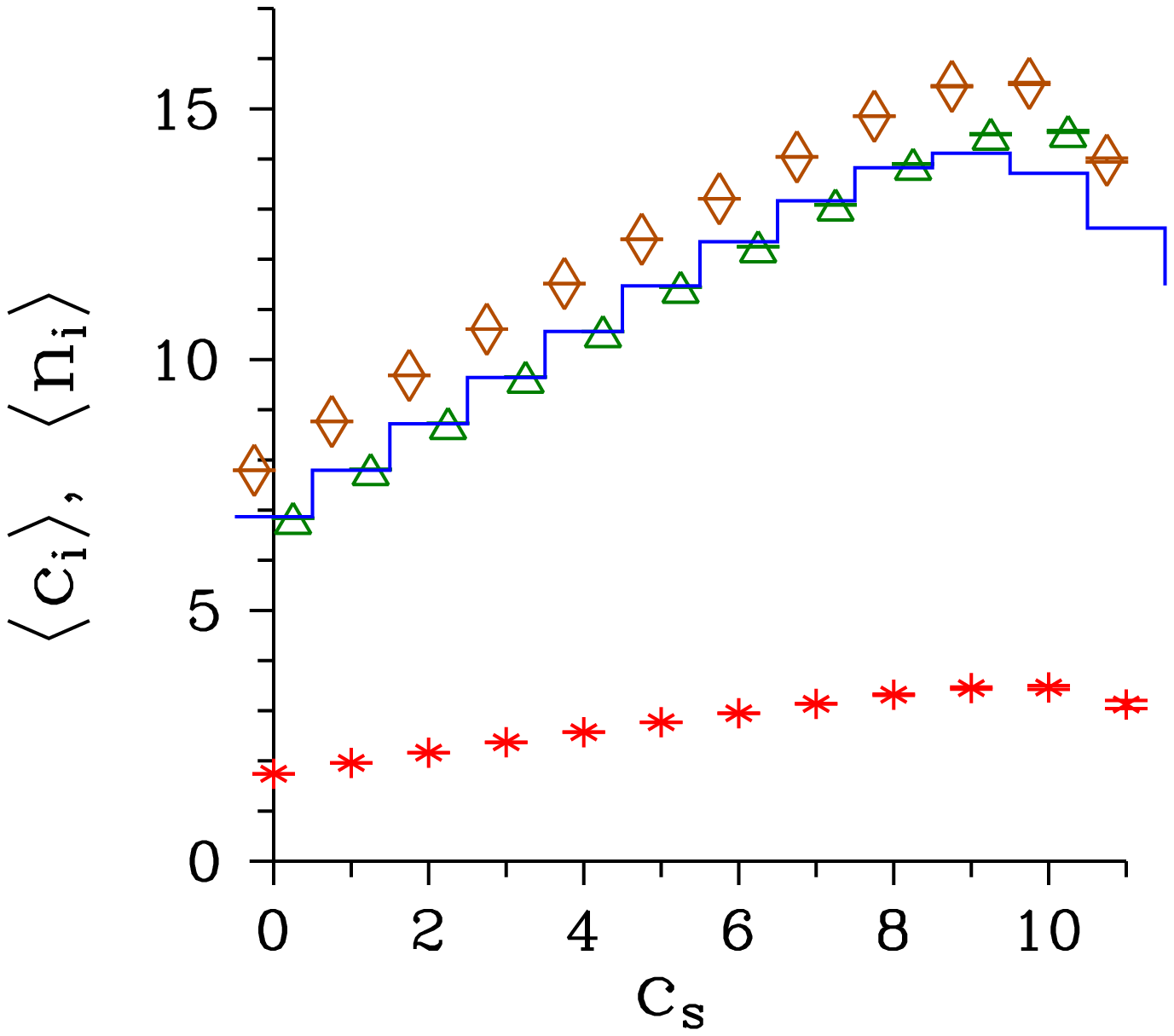}}}
 \centerline{\small (a) \hspace{0.2\textwidth} (b)}
 \caption{(a) Signal photocount histogram $ f_{\rm s} $
  and its theoretical prediction
  $ f_{\rm s}^{\rm theo} $ and (b) mean photocount (photon) number
  $ \langle c_{\rm i} \rangle $ ($ \langle n_{\rm i} \rangle $) of the post-selected idler
  fields as they depend on signal photocount number $ c_{\rm s} $. Data for
  experimental photocount distributions (red $ \ast $) and photon-number
  distributions determined by MLA (green $ \triangle $),
  GTWB (blue solid curve) and modifying the intensity moments
  (brown $ \diamond $) are plotted. Experimental errors are smaller than the used
  symbols.}
\end{figure}

\emph{Criteria I and II:} Parameters $ r_W^{(k)} $ quantifying $
k$-th-order non-classicalities via the 'theoretical' intensity
moments and plotted in Figs.~3(a,c,e,g) show that the
post-selected idler fields conditioned by the detected signal
photocount numbers $ c_{\rm s} $ in the range $ \langle 3,7
\rangle $ are nonclassical in the second and the third orders.
Moreover the post-selected idler fields in the range $ \langle
5,7\rangle $ are nonclassical in the fourth and the fifth orders.
The comparison of intensity parameters $ r_W^{(k)} $ with the
corresponding 'experimental' photon-number parameters $ r_n^{(k)}
$ based on the graphs in Fig.~3 reveals accordance in the
occurrence of non-classicality of different orders indicated by
both kinds of parameters for the measured photocount as well as
the reconstructed photon-number quantities. The graphs in Fig.~3
also show that greater negative values of parameters $ r_W^{(k)} $
and $ r_n^{(k)} $ are systematically reached for the reconstructed
photon-number distributions (by MLA and GTWB) in comparison with
those arising in the photocount distributions. This is due to
partial elimination of the noise by the reconstruction.
\begin{figure}  % fig. 3
 \centerline{\resizebox{0.45\hsize}{!}{\includegraphics{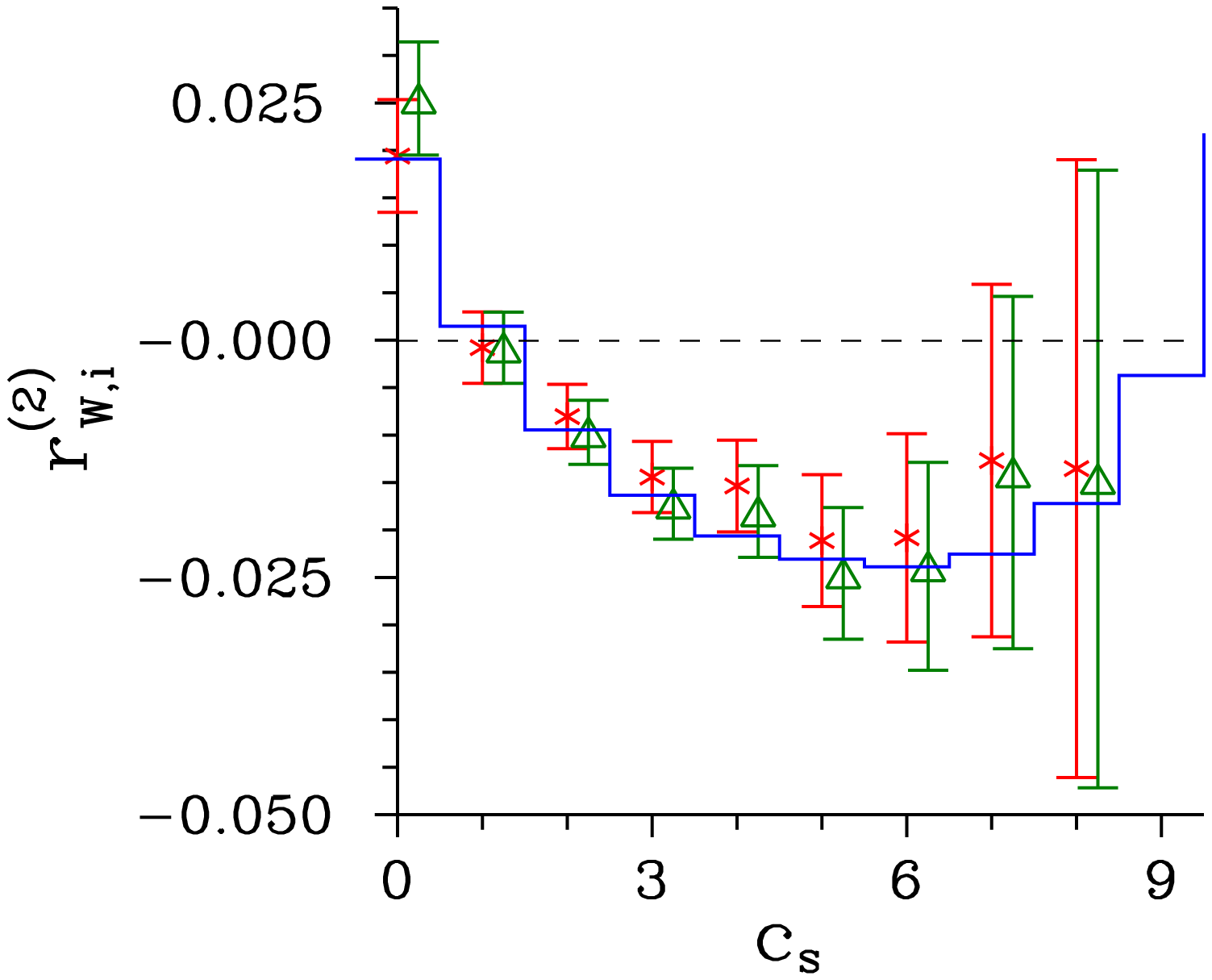}}
  \hspace{2mm} \resizebox{0.45\hsize}{!}{\includegraphics{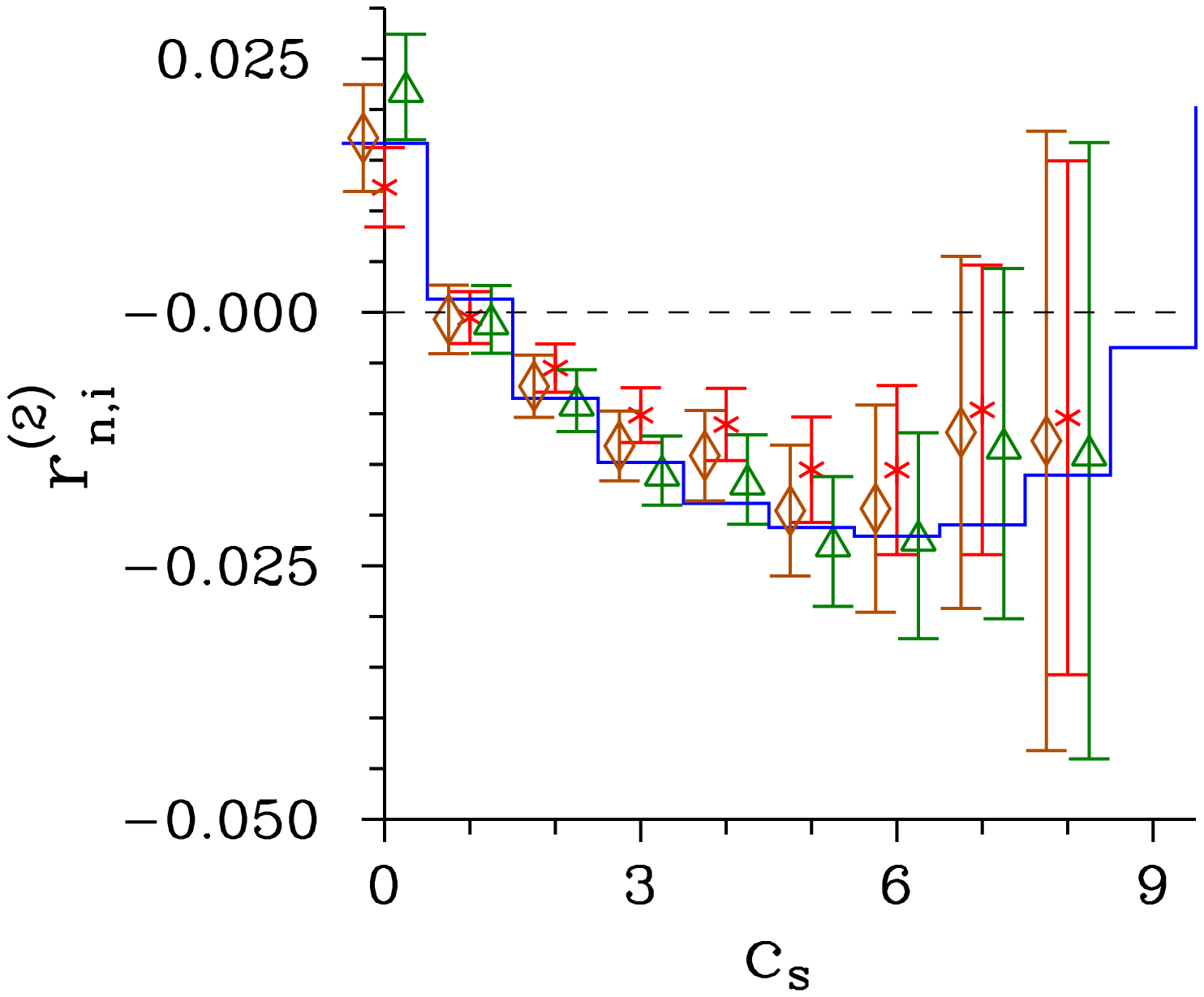}}}
 \centerline{\small (a) \hspace{0.2\textwidth} (b)}
 \vspace{1mm}
 \centerline{\resizebox{0.45\hsize}{!}{\includegraphics{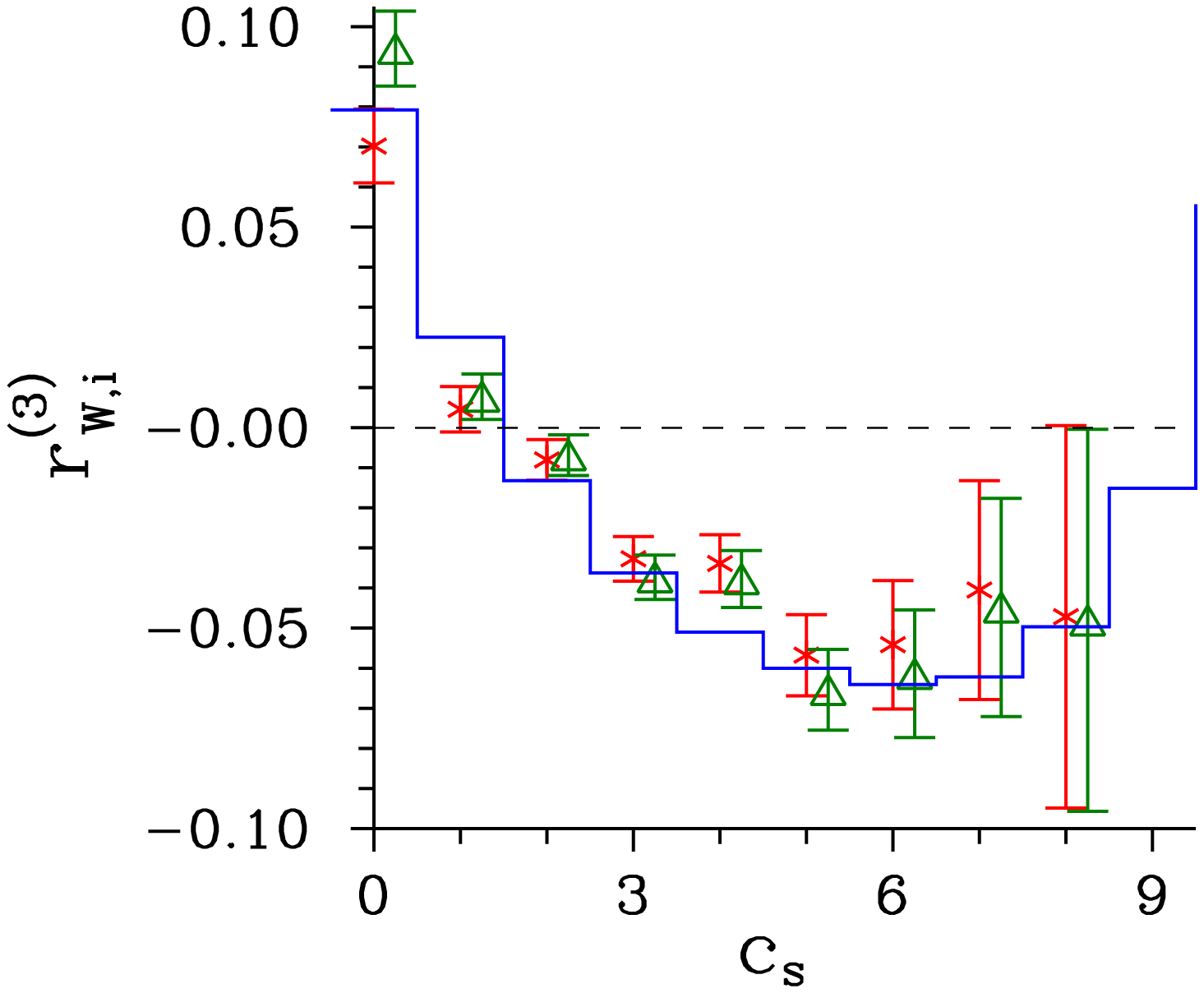}}
  \hspace{2mm} \resizebox{0.45\hsize}{!}{\includegraphics{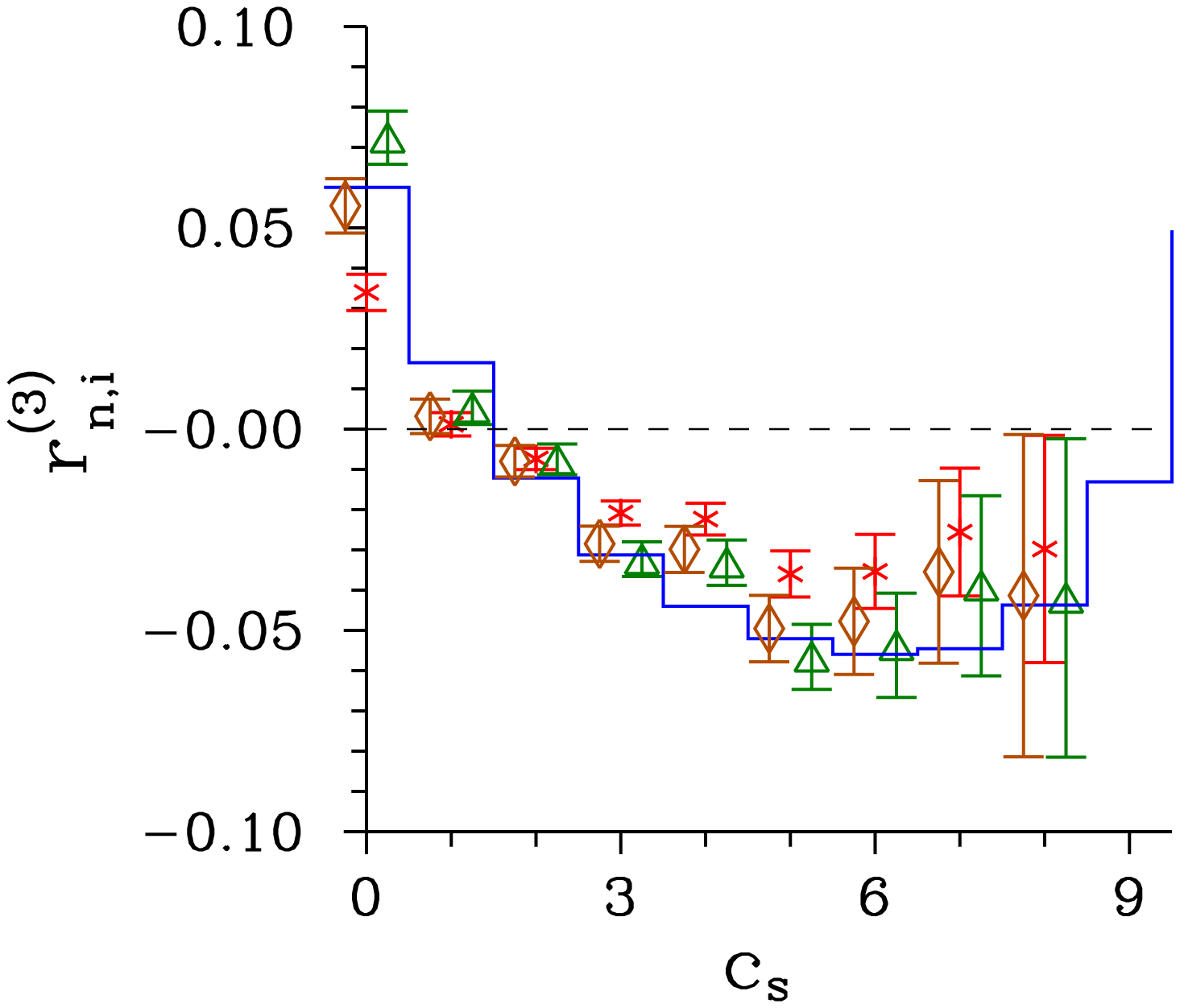}}}
 \centerline{\small (c) \hspace{0.2\textwidth} (d)}
 \vspace{1mm}
 \centerline{\resizebox{0.45\hsize}{!}{\includegraphics{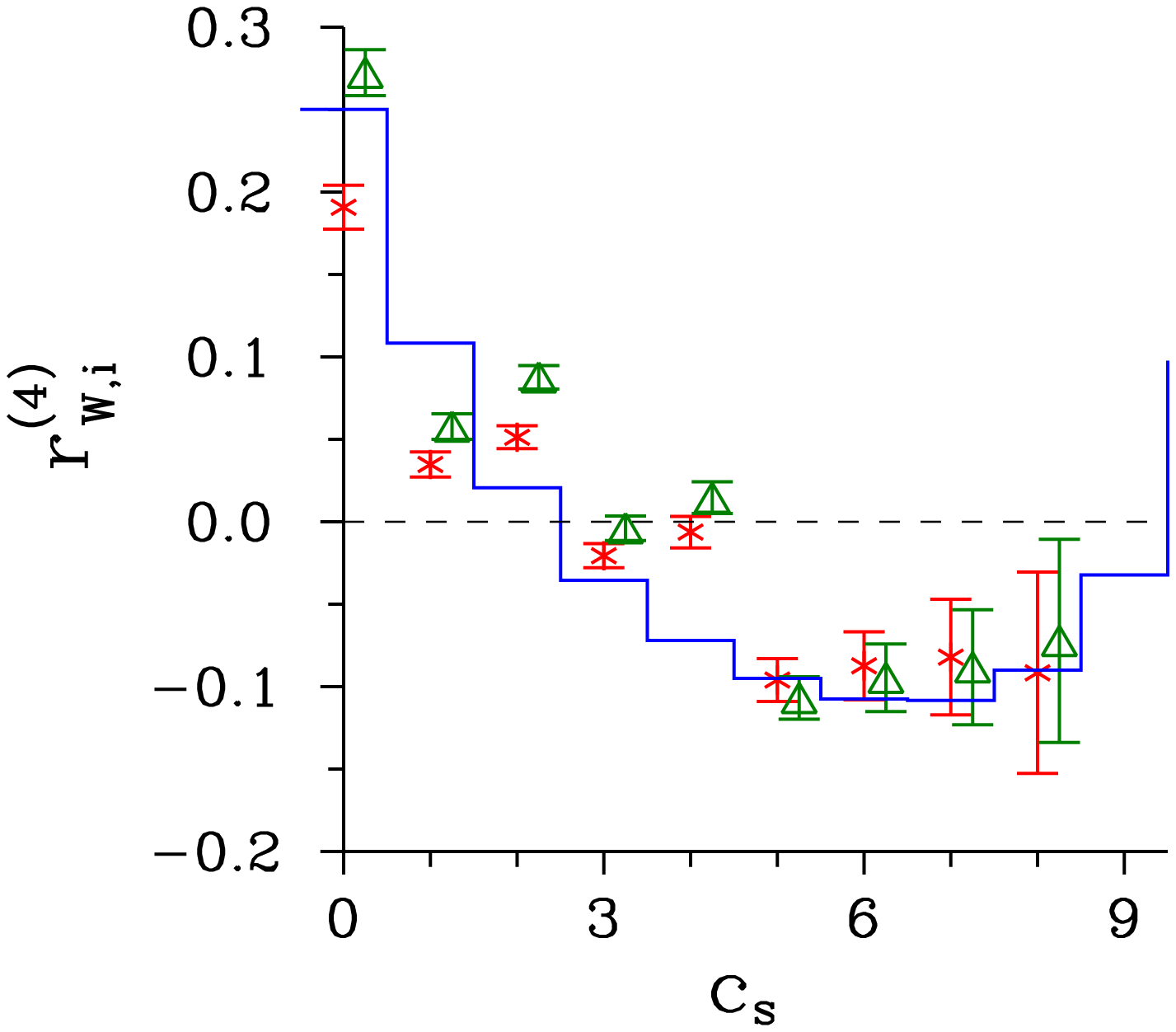}}
  \hspace{2mm} \resizebox{0.45\hsize}{!}{\includegraphics{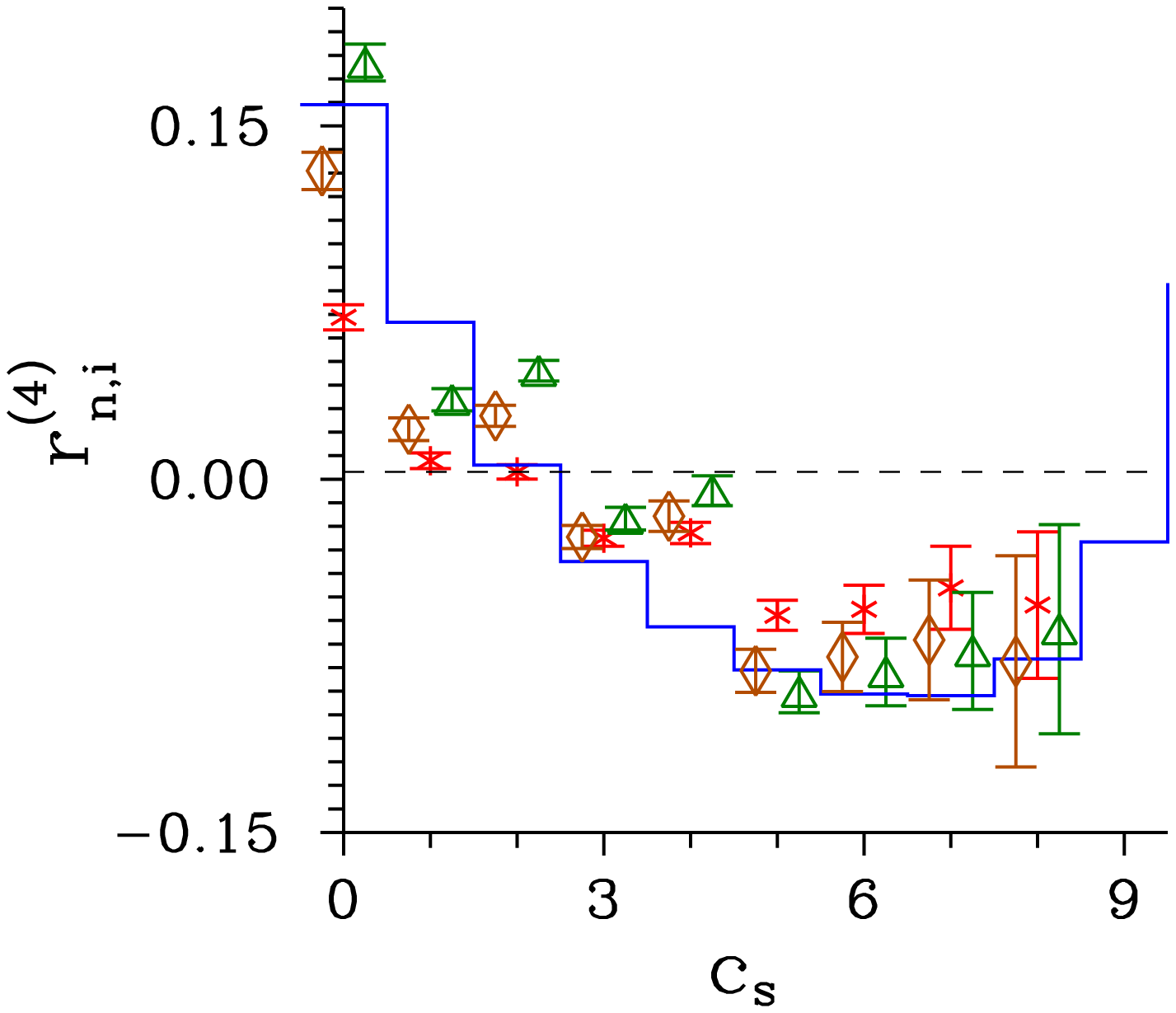}}}
 \centerline{\small (e) \hspace{0.2\textwidth} (f)}
 \vspace{1mm}
 \centerline{\resizebox{0.45\hsize}{!}{\includegraphics{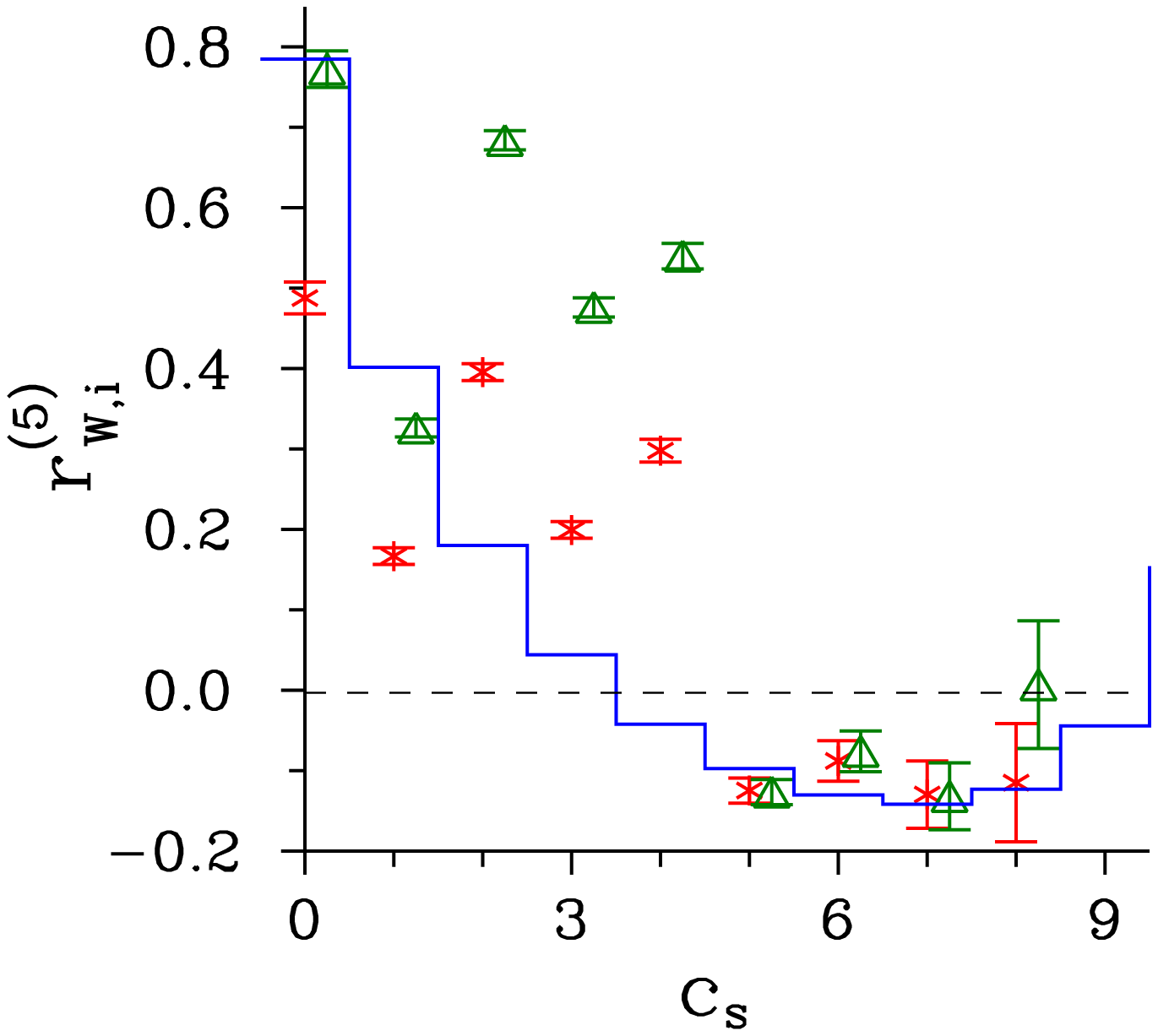}}
  \hspace{2mm} \resizebox{0.45\hsize}{!}{\includegraphics{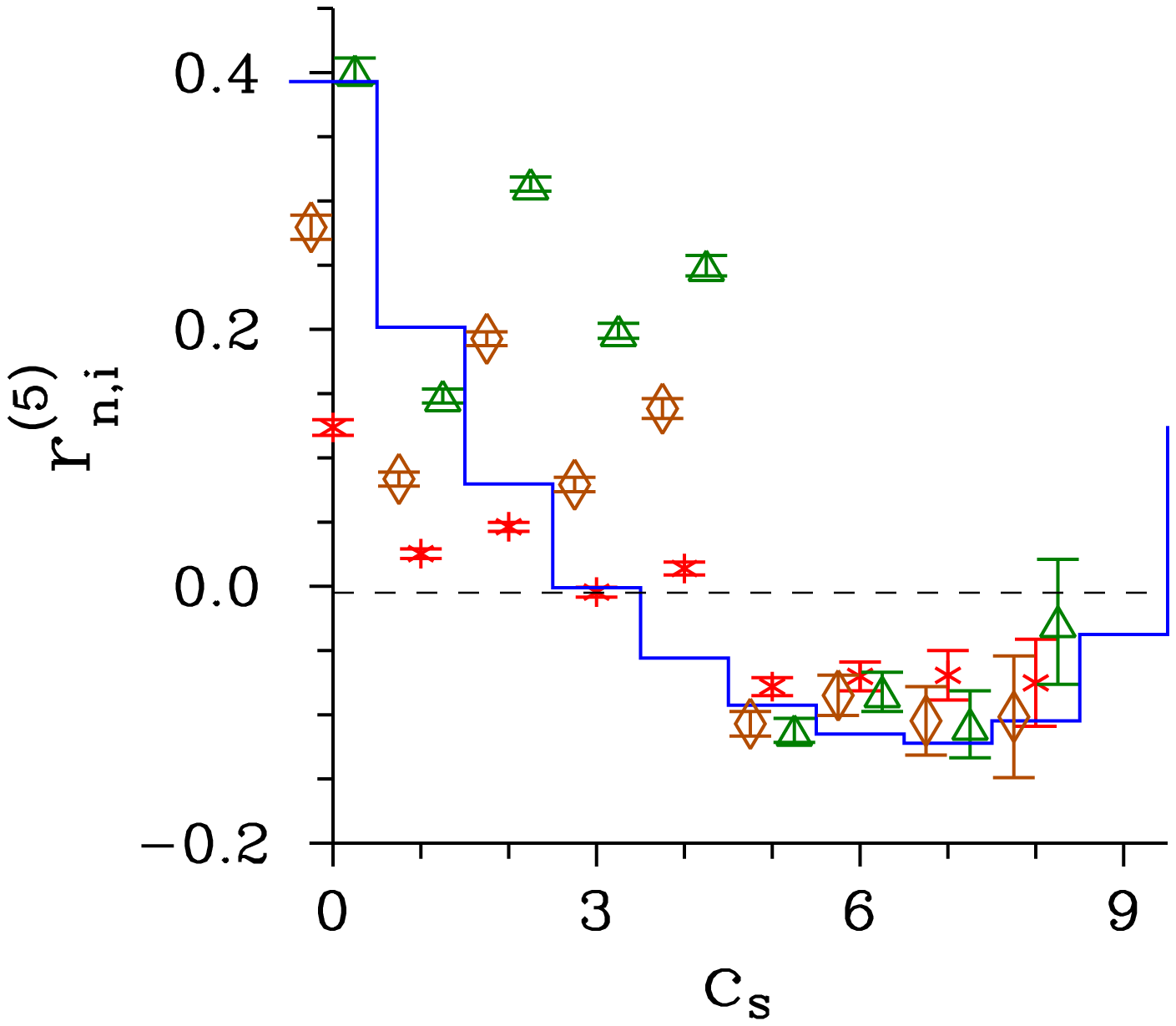}}}
 \centerline{\small (g) \hspace{0.2\textwidth} (h)}
 \caption{Parameters $ r_{W,{\rm i}}^{(k)} $ and $ r_{n,{\rm i}}^{(k)} $ of
  the post-selected idler fields
  for $ k= 2 $ (a,b), 3 (c,d), 4 (e,f), and 5 (g,h) as they depend
  on signal photocount number $ c_{\rm s} $. The used symbols and curves
  are described in the caption to Fig.~2.}
\end{figure}

The non-classicalities of different orders are mutually compared
in Fig.~4 via their non-classicality depths $ \tau^{(k)} $ defined
in Eq.~(\ref{9}). For the generated states, the greater the
non-classicality order $ k $ is the smaller the values $
\tau^{(k)} $ of the corresponding non-classicality depths are
observed and so the weaker the resistance of the non-classicality
against the external noise is. As the directly measured photocount
distributions give roughly 4-times lower intensities than the
reconstructed photon-number distributions, the obtained values of
non-classicality depths $ \tau^{(k)} $ are naturally smaller for
photocounts compared to photon numbers.
\begin{figure}  % fig. 4
 \centerline{\resizebox{0.45\hsize}{!}{\includegraphics{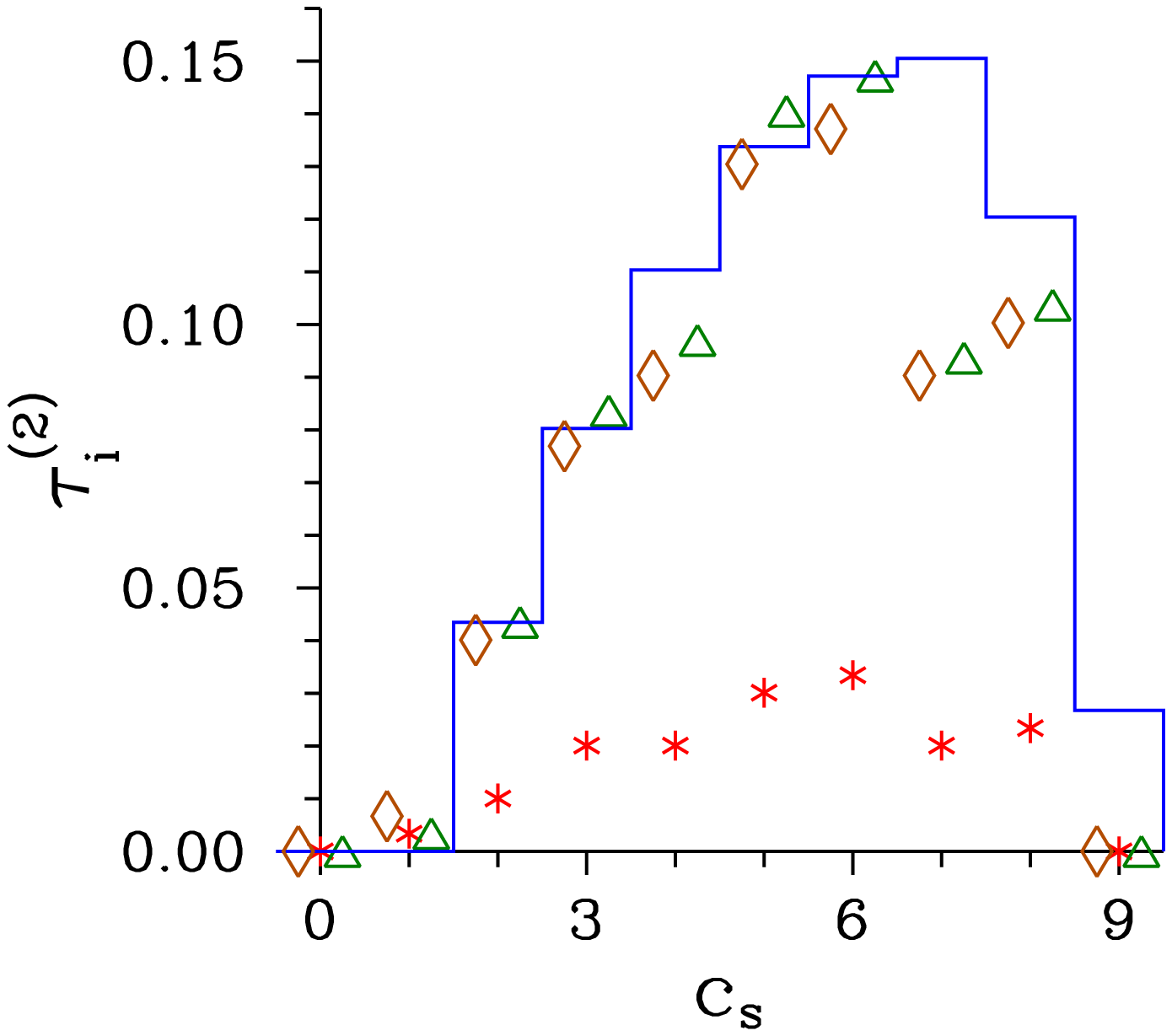}}
  \hspace{2mm} \resizebox{0.45\hsize}{!}{\includegraphics{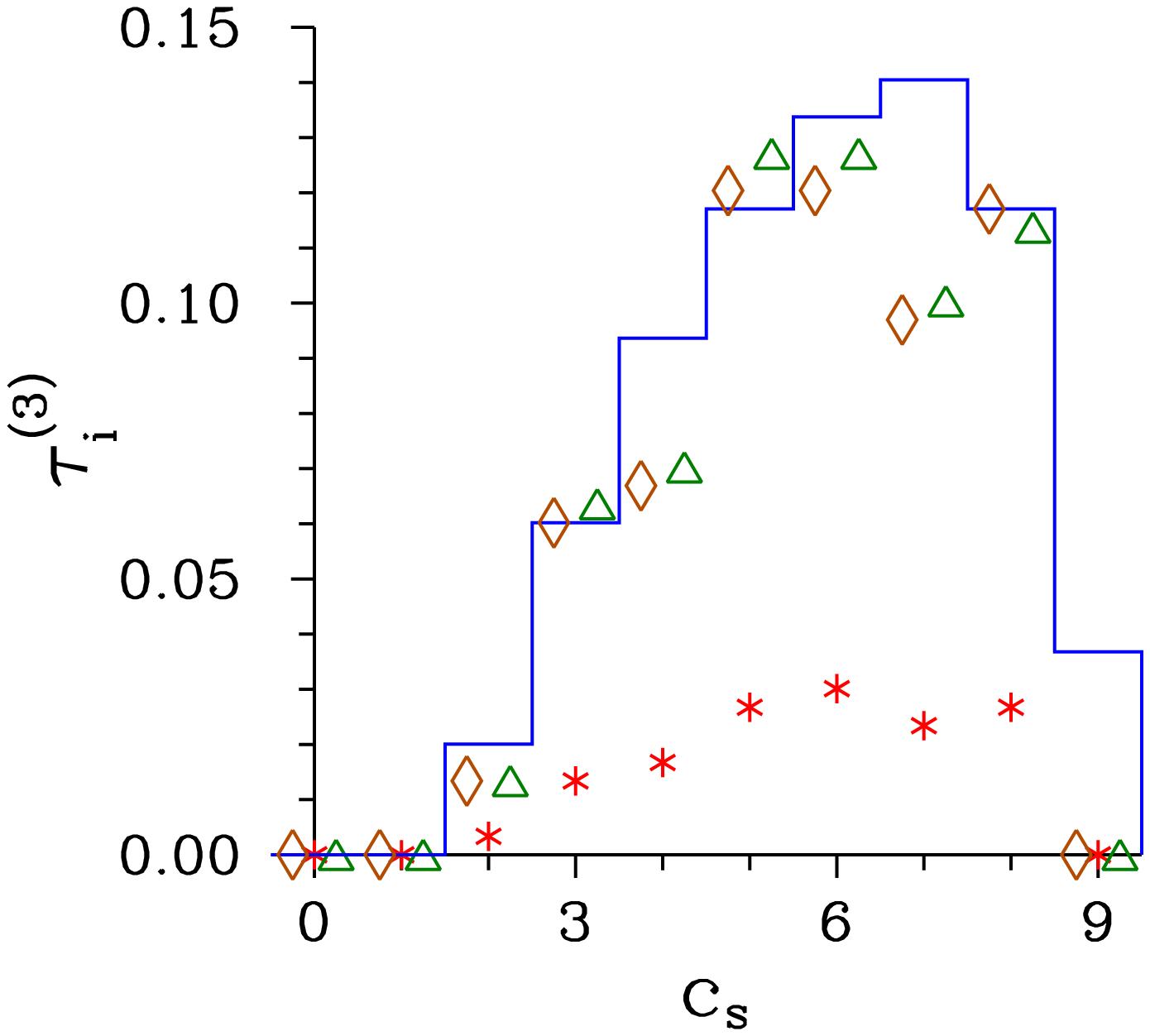}}}
 \centerline{\small (a) \hspace{0.2\textwidth} (b)}
 \vspace{1mm}
 \centerline{\resizebox{0.45\hsize}{!}{\includegraphics{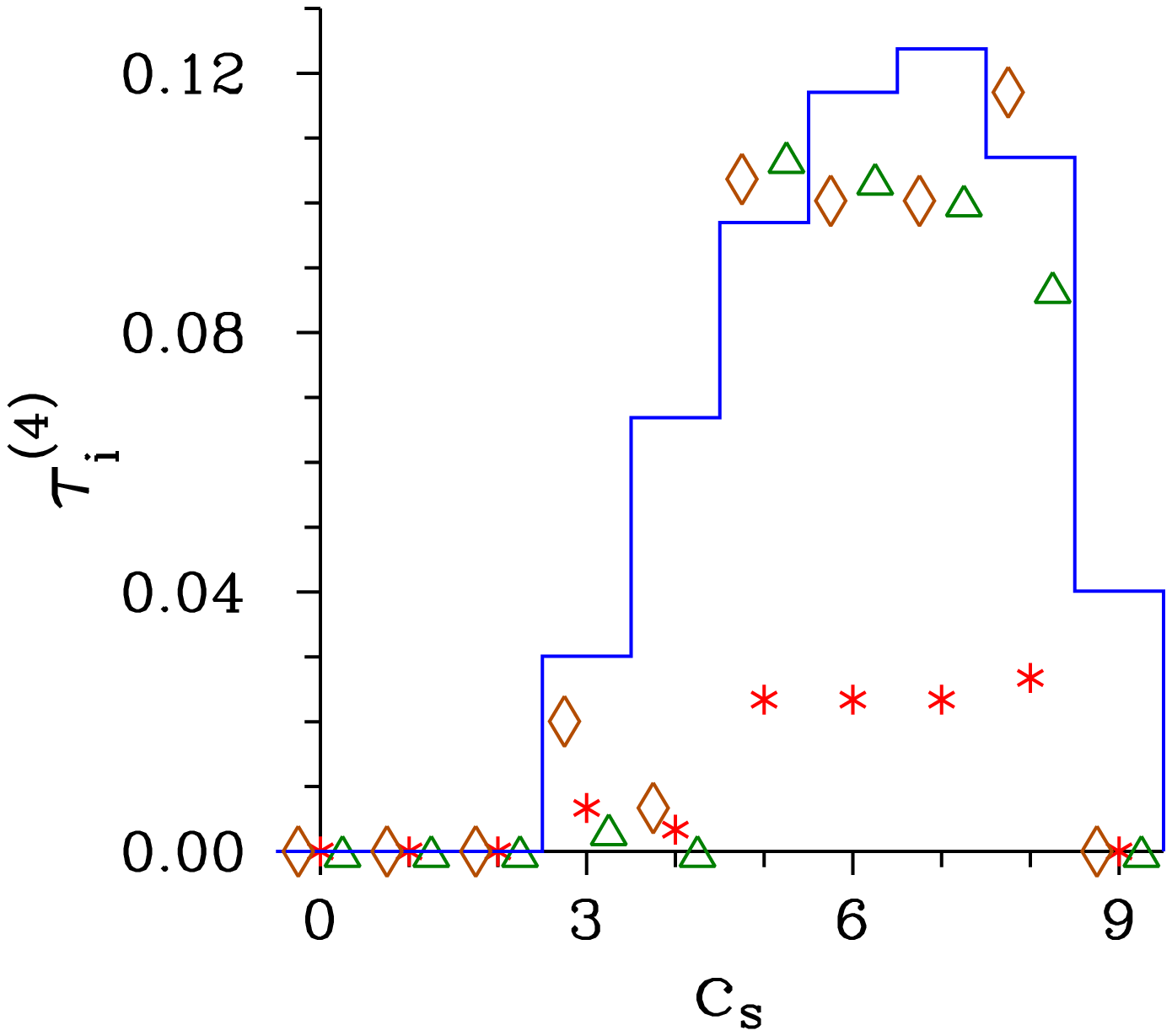}}
  \hspace{2mm} \resizebox{0.45\hsize}{!}{\includegraphics{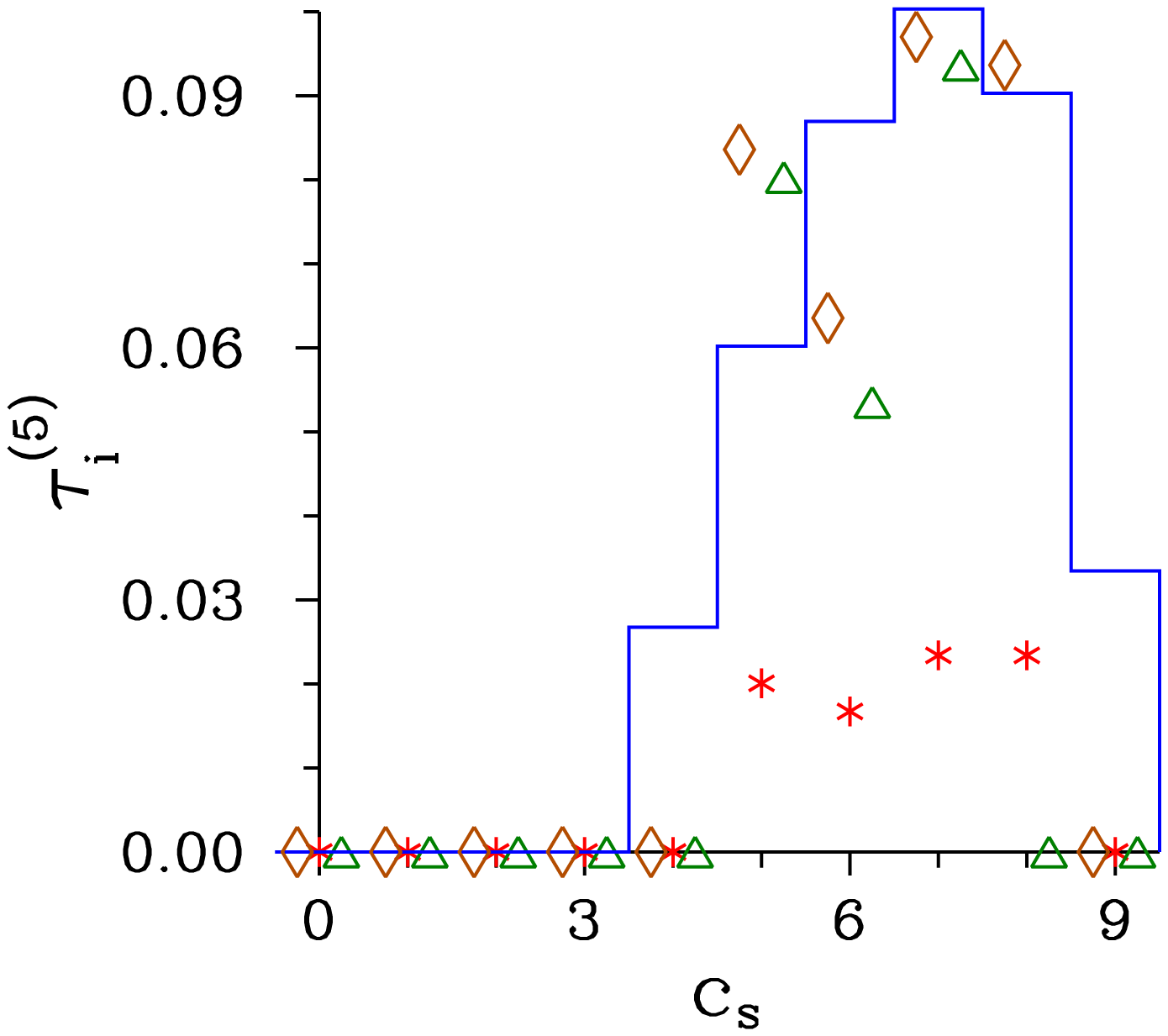}}}
 \centerline{\small (c) \hspace{0.2\textwidth} (d)}
 \caption{Non-classicality depths $ \tau_{\rm i}^{(k)} $ of
  the post-selected idler
  fields for $ k= 2 $ (a), 3 (b), 4 (c), and 5 (d) as they depend
  on signal photocount number $ c_{\rm s} $. The used symbols and curves
  are described in the caption to Fig.~2.}
\end{figure}

\emph{Criteria III and IV:} Excluding the second-order
non-classicalities for which the parameters $ r_{\Delta W}^{(2)} $
and $ r_{\Delta n}^{(2)} $ accord with the above analyzed
parameters $ r_W^{(2)} $ and $ r_n^{(2)} $, the ability of both
experimental and reconstructed moments of intensity and photocount
(photon-number) fluctuations to reveal higher-order
non-classicalities is qualitatively worse than that of parameters
$ r_W^{(k)} $ and $ r_n^{(k)} $. This is due to large experimental
errors and the reasons discussed below Eq.~(\ref{6}). As shown in
Fig.~5, none of the non-classicality identifiers $ r_{\Delta
W}^{(4)} $, $ r_{\Delta c}^{(3)} $ and $ r_{\Delta c}^{(4)} $
reveals the post-selected idler fields as non-classical using the
experimental photocounts and their errors.
\begin{figure}  % fig. 5
 \centerline{\resizebox{0.45\hsize}{!}{\includegraphics{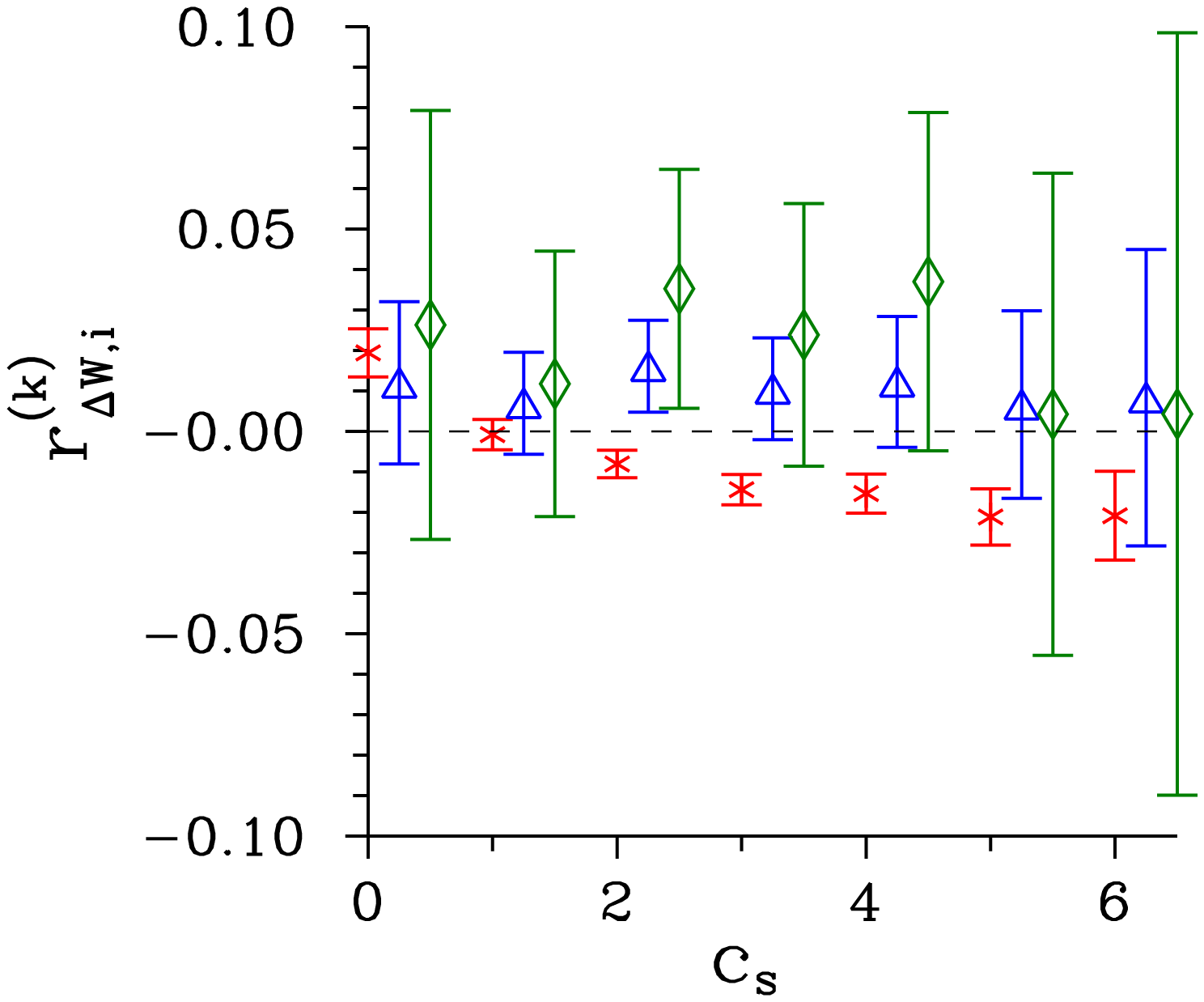}}
  \hspace{2mm} \resizebox{0.45\hsize}{!}{\includegraphics{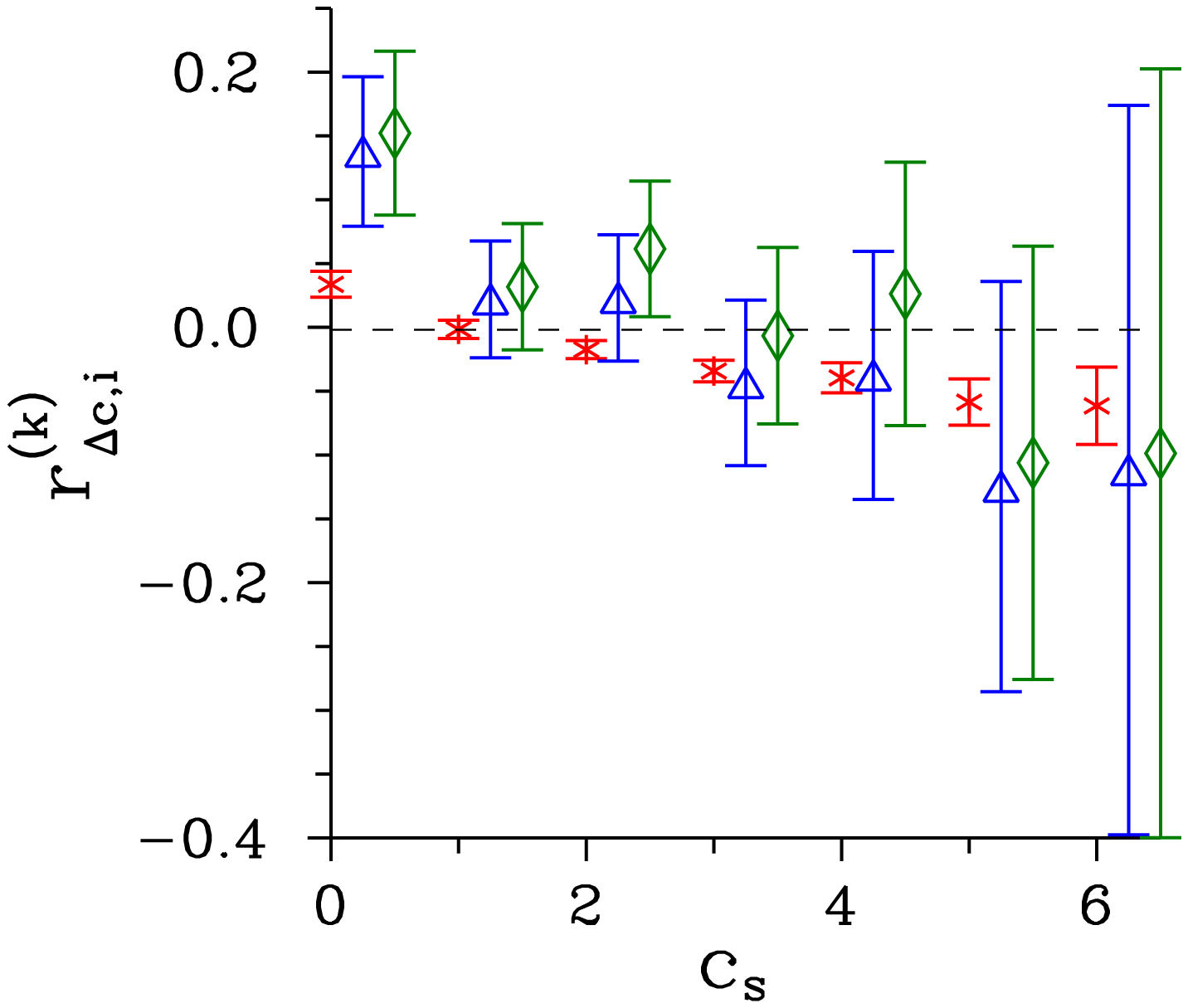}}}
 \centerline{\small (a) \hspace{0.2\textwidth} (b)}
 \caption{Parameters $ r_{\Delta W,{\rm i}}^{(k)} $ (a) and $ r_{\Delta c,{\rm i}}^{(k)} $ (b) of the
  post-selected photocount idler fields for $ k= 2 $ (red $ \ast $), 3 (blue $ \triangle $) and 4 (green $ \diamond $) as they depend
  on signal photocount number $ c_{\rm s} $.}
\end{figure}

\emph{Criterion V:} From the point of view of experimental errors,
the best results are found for the parameters $ r_p^{(k)} $
defined in Eq.~(\ref{8}) and involving the elements $ f_{\rm
i}(c_{\rm i};c_{\rm s}) $ of conditional idler photocount
histograms, that represent a certain discrete transform of the
moments $ \langle c_{\rm i}^l \rangle $ for $ l=0,1,\ldots, \infty
$. The graphs presented in Fig.~6 document that we can recognize,
within the experimental errors, the non-classicality up to the
ninth order for idler fields post-selected by the signal
photocount numbers $ c_{\rm s} $ in the range $ \langle 2,7\rangle
$. Also, the idler field post-selected by the detection of $
c_{\rm s} =1 $ signal photocount is newly identified as
nonclassical, even up to the fifth order. On the other hand, the
reconstruction procedures performed for detection efficiency $
\eta_{\rm i} \approx 0.2 $ considerably amplify the fields and
thus introduce larger errors. This disqualifies the application of
parameters $ r_p^{(k)} $ to the reconstructed fields.
\begin{figure}  % fig. 6
 \centerline{\resizebox{0.45\hsize}{!}{\includegraphics{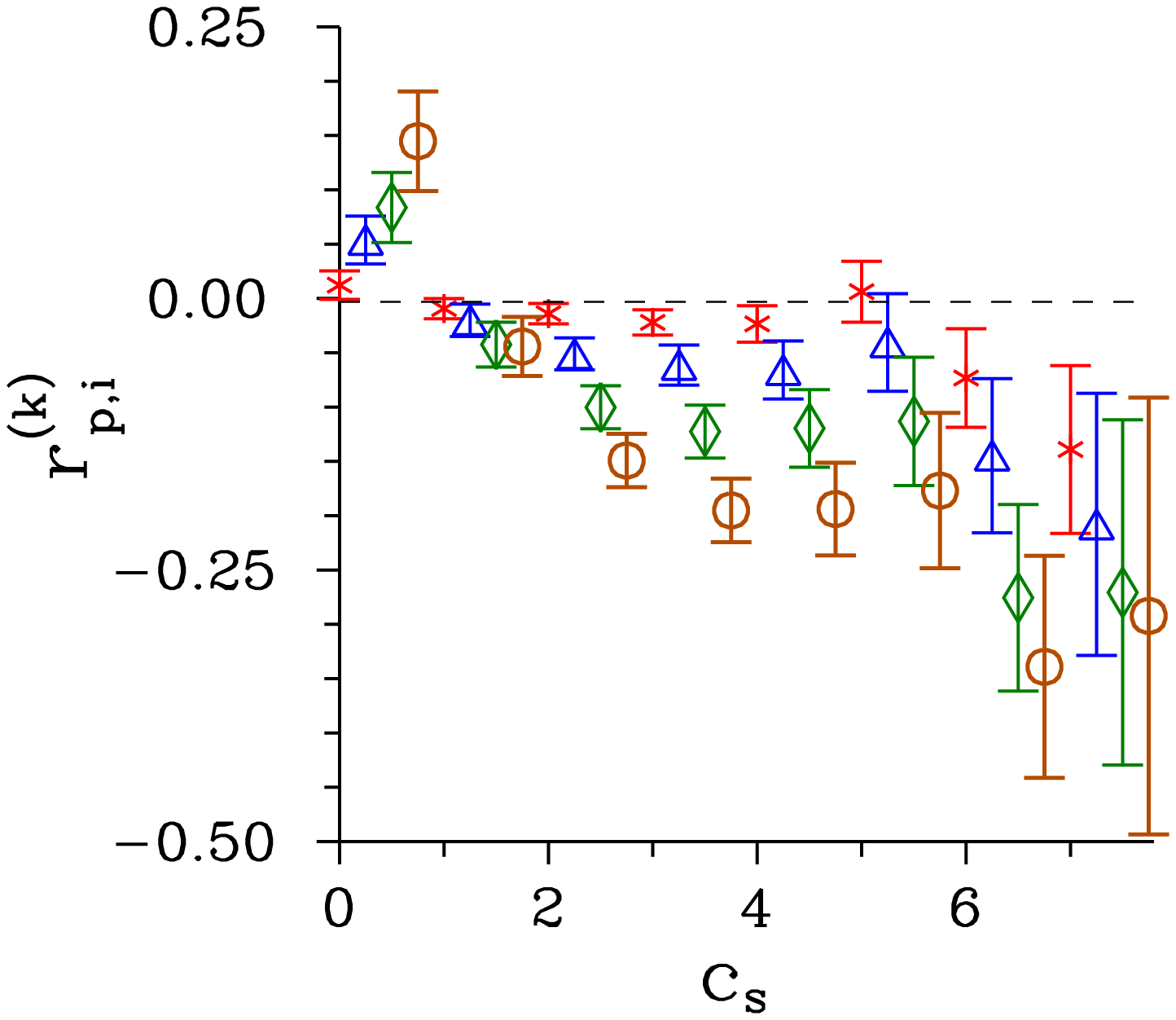}}
  \hspace{2mm} \resizebox{0.45\hsize}{!}{\includegraphics{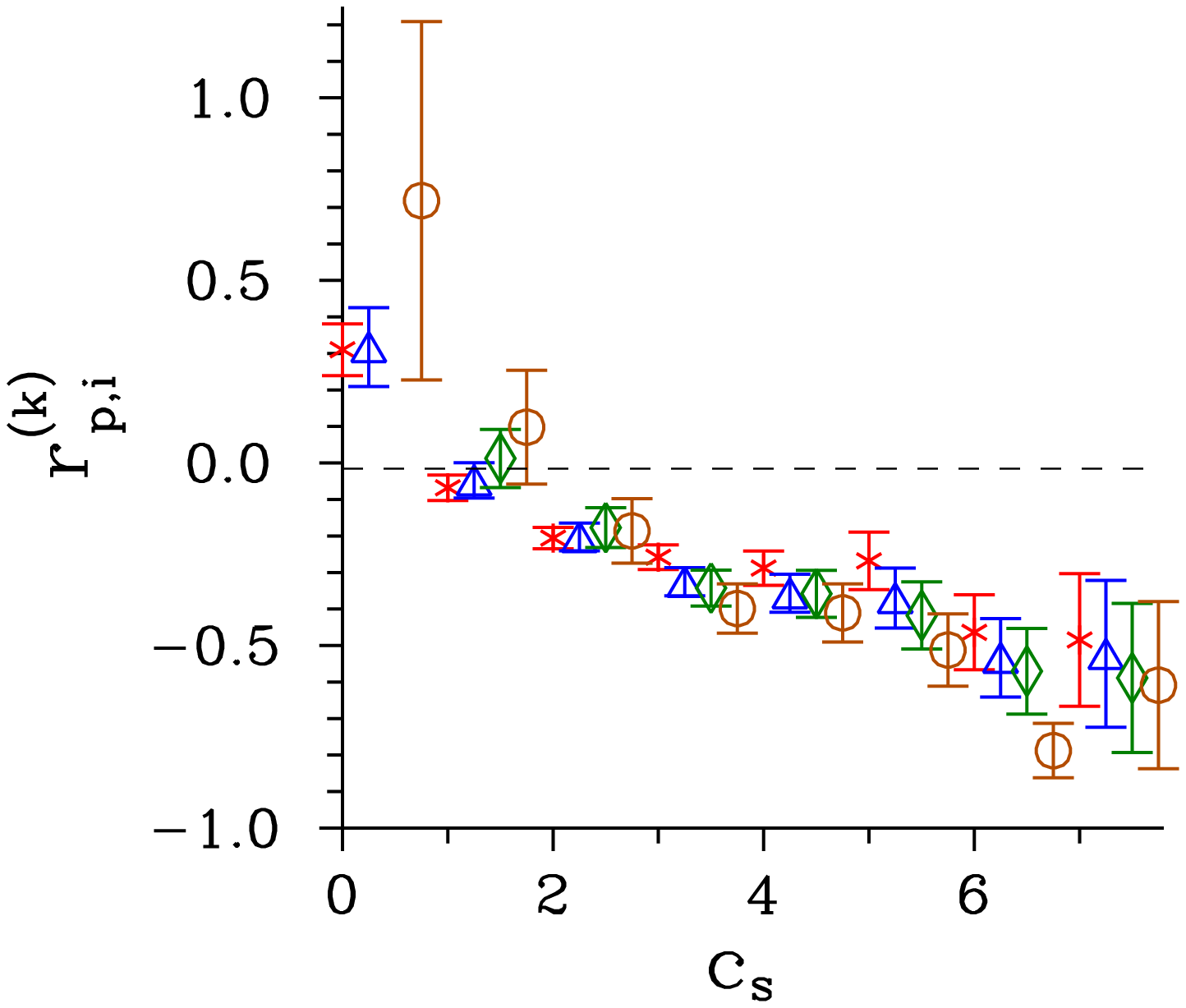}}}
 \centerline{\small (a) \hspace{0.2\textwidth} (b)}
 \caption{Parameters $ r_{p,{\rm i}}^{(k)} $ of the post-selected idler
  photocount histograms for (a) $ k= 2 $ (red $ \ast $), 3 (blue $ \triangle $), 4
  (green $ \diamond $), and 5 (brown $ \circ $) and (b) $ k= 6 $ (red $ \ast $),
  7 (blue $ \triangle $), 8 (green $ \diamond $), and 9 (brown $ \circ $)
  as they depend on signal photocount number $ c_{\rm s} $.}
\end{figure}

At last, we present, as a typical example, the distributions
characterizing the idler fields post-selected by the detection of
$ c_{\rm s} = 5 $ signal photocounts. The experimental idler
photocount histogram $ f_{\rm i}(c_{\rm i}) $ is compared with the
corresponding Poissonian distribution in Fig.~7(a). Similarly, the
post-selected idler photon-number distributions $ p_{\rm
c,i}(n_{\rm i}) $ reached by MLA and GTWB are compared with the
appropriate Poissonian distribution in Fig.~7(b). Whereas the
photocount histogram is very close to its Poissonian counterpart,
the difference between the reconstructed photon-number
distributions and the Poissonian distribution is well recognized.
\begin{figure}  % fig. 7
 \centerline{\resizebox{0.45\hsize}{!}{\includegraphics{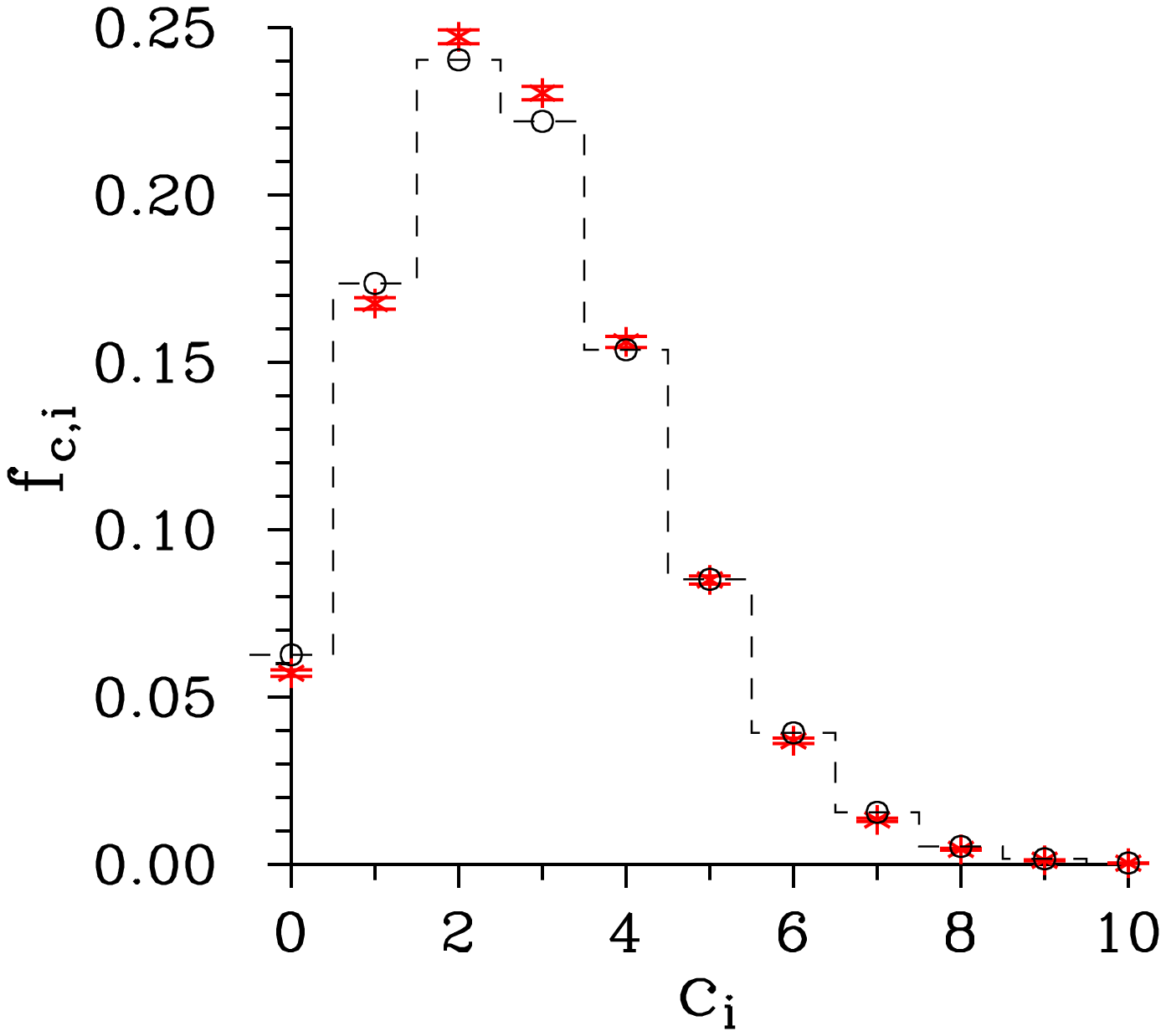}}
  \hspace{2mm} \resizebox{0.45\hsize}{!}{\includegraphics{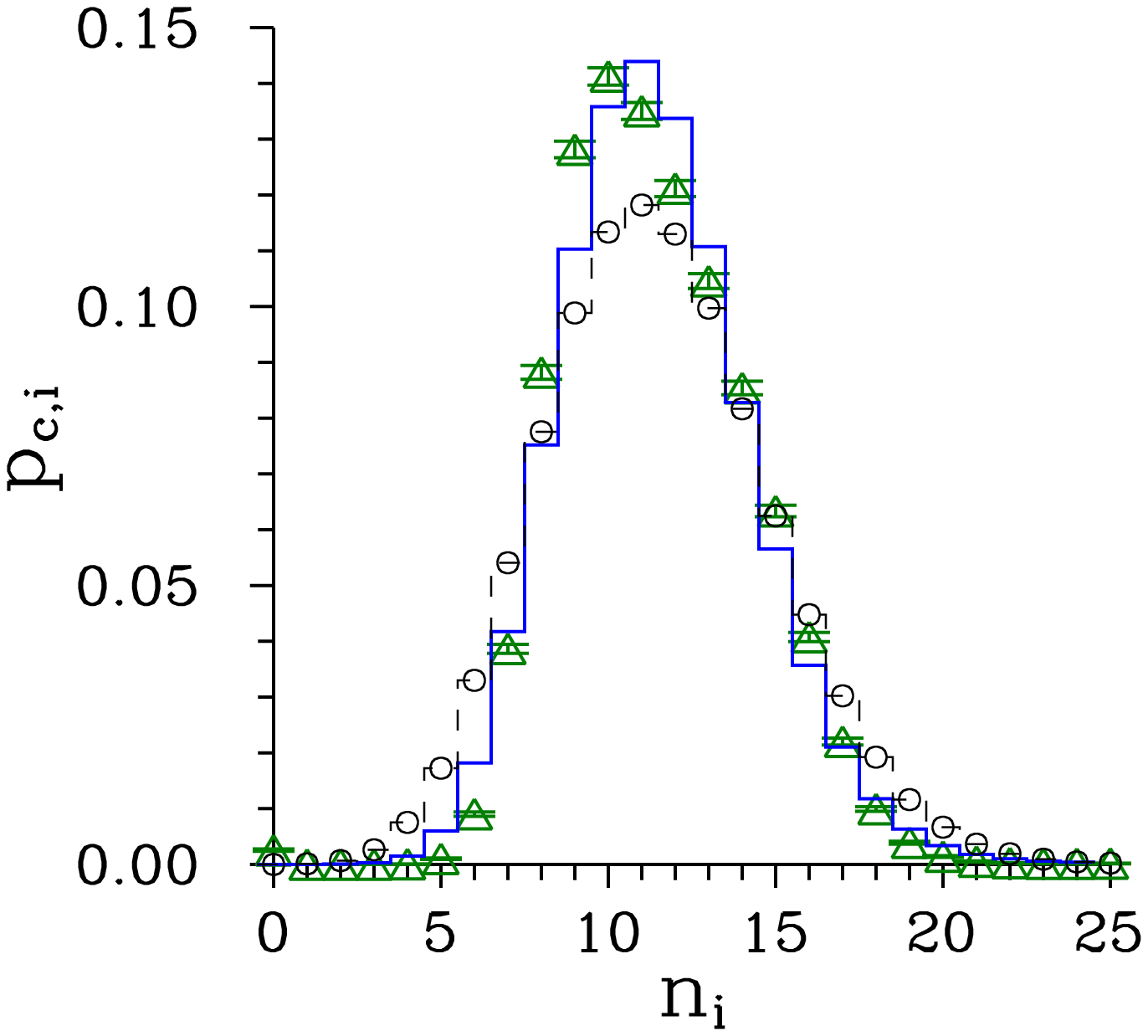}}}
 \centerline{\small (a) \hspace{0.2\textwidth} (b)}
 \caption{(a) Idler photocount histogram $ f_{\rm c, i}(c_{\rm i})\equiv f_{\rm i}
  (c_{\rm i};c_{\rm s}) $ and (b)
  idler photon-number distributions $ p_{\rm c,i}(n_{\rm i}) $
  post-selected by detection of $ c_{\rm s} = 5 $ signal photocounts. The used symbols and blue
  curve are described in the caption to Fig.~2. For comparison,
  the corresponding Poissonian distributions are drawn by dashed
  curves with $ \circ $.}
\end{figure}

The signs of intensity parameters $ r_{W,s}^{(k)} $ defined in
terms of $ s $-ordered intensity moments qualitatively influence
the shape of quasi-distribution $ \tilde P_{\rm c,i}(W_{\rm i};s)
$ of integrated intensity. If the parameters $ r_{W,s}^{(k)} $ are
negative, the difference $ \Delta \tilde P_{\rm c,i} $ defined in
Eq.~(\ref{11}) attains negative values in the regions where they
are not compensated by positive values of the Poissonian
distribution $ \tilde P_{\rm Pois} $ and so the resultant
quasi-distribution $ \tilde P_{\rm c,i}(W_{\rm i};s) $ is
nonclassical due to its negative values. This occurs for the
ordering parameter $ s $ greater than $ s_{\rm th}^{(2)} $, as
demonstrated in Fig.~8(a) for $ s=0.9 $. On the other hand,
positive parameters $ r_{W,s}^{(k)} $ observed for $ s < s_{\rm
th}^{(2)} $ cause 'redistribution' of the classical probability
densities of $ \tilde P_{\rm c,i}(W_{\rm i};s) $ such that greater
values occur for small intensities $ W $ and the central peak
lowers [see Fig.~8(b)].
\begin{figure}  % fig. 8
 \centerline{\resizebox{0.45\hsize}{!}{\includegraphics{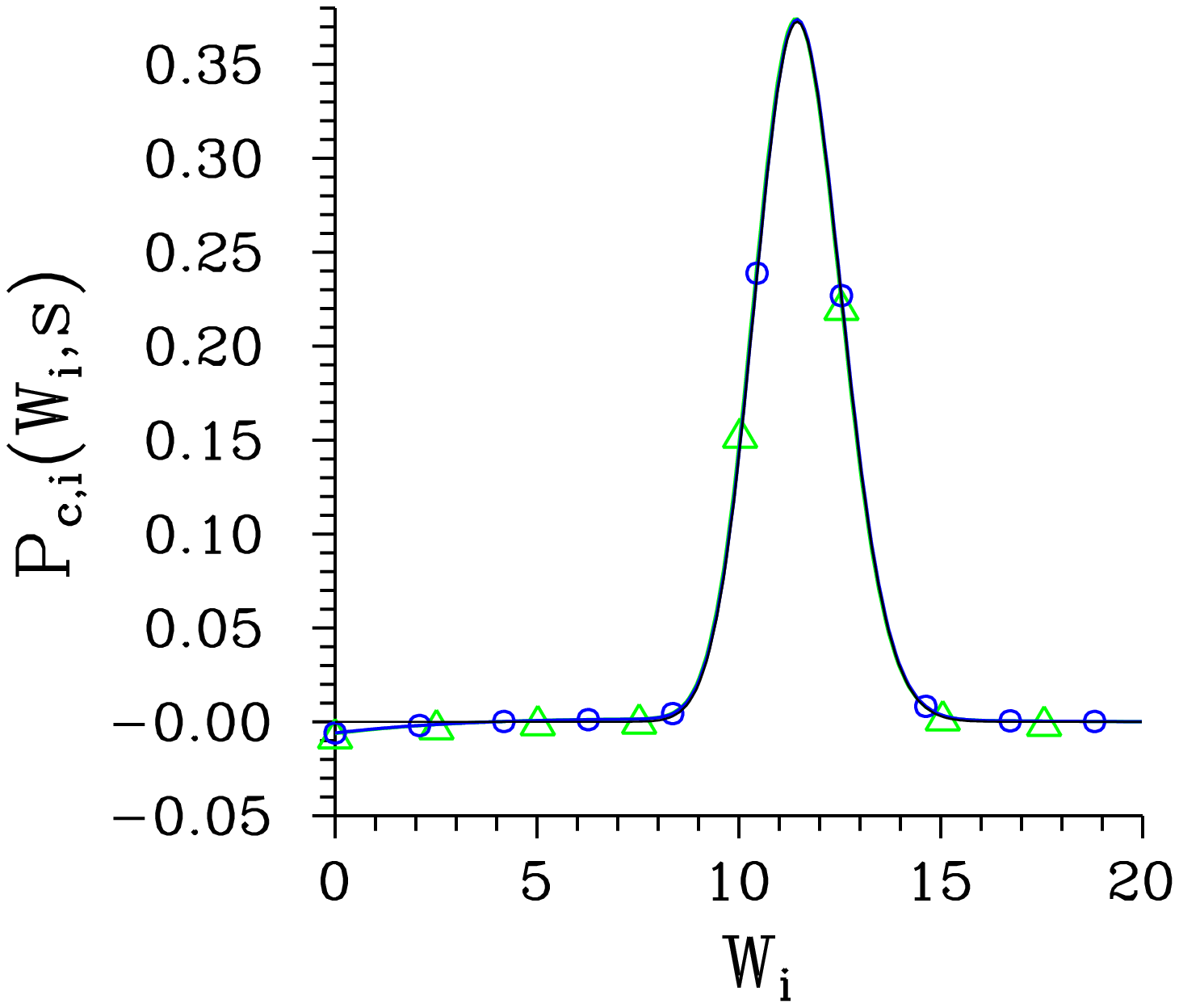}}
  \hspace{2mm} \resizebox{0.45\hsize}{!}{\includegraphics{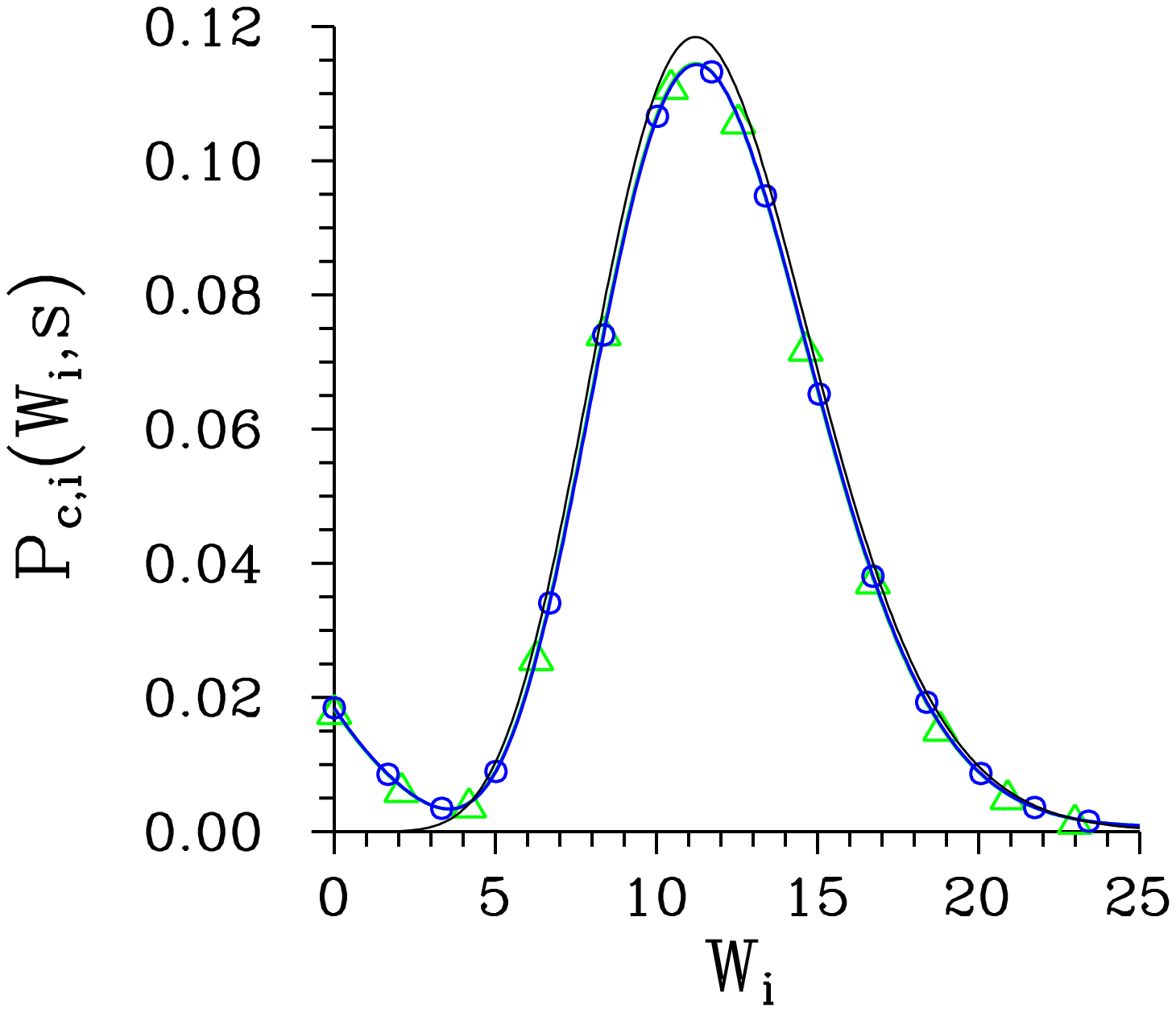}}}
 \centerline{\small (a) \hspace{0.2\textwidth} (b)}
 \caption{Quasi-distributions $ \tilde P_{\rm c,i}(W_{\rm i};s) $ of integrated
  intensity derived from those of the
  Poissonian field via Eq.~(\ref{11}) \cite{Perina1991} for (a) $ s=0.9 $ and
  (b) $ s=0 $ for the idler field post-selected by detection of $
  c_{\rm s} = 5 $ signal photocounts obtained by MLA (green curves with $ \triangle $) and
  GTWB (blue curves with $ \circ $); $ \tau^{(2)}_{\rm i} \approx 0.14 $.
  Distributions $ \tilde P_{\rm Pois} $ of the corresponding
  Poissonian fields are drawn by plain curves for comparison.}
\end{figure}

In conclusion, we have shown that higher-order sub-Poissonian-like
criteria based on intensity and photocount (photon-number) moments
are suitable and comparably strong for revealing higher-order
non-classicalities. Contrary to this, the criteria exploiting
moments of intensity and photocount (photon-number) fluctuations
have been found not very useful owing to their sensitivity to
experimental errors. From the point of view of experimental
errors, the criteria based on the elements of photocount
distributions have been identified as the most powerful allowing
us to experimentally reach even the ninth-order non-classicality.

\section*{Acknowledgments} The authors thank M. Hamar for his help
with the experiment. J. P. Jr acknowledges the discussion with
J.~Pe\v{r}ina concerning the reconstruction of quasi-distributions
of integrated intensity. The authors were supported by the GA
\v{C}R (project 15-08971S) and M\v{S}MT \v{C}R (project LO1305).

% \bibliography{perina}

\begin{thebibliography}{10}

\bibitem{Luks1988}
A.~Luk\v{s}, V.~Pe\v{r}inov\'{a}, and J.~Pe\v{r}ina.
\newblock Principal squeezing of vacuum fluctuations.
\newblock {\em Opt. Commun.}, 67:149---151, 1988.

\bibitem{Davidovich1996}
L.~Davidovich.
\newblock {Sub-Poissonian} processes in quantum optics.
\newblock {\em Rev. Mod. Phys.}, 68:127---173, 1996.

\bibitem{Kimble1977}
H.~J. Kimble, M.~Dagenais, and L.~Mandel.
\newblock Photon antibunching in resonance fluorescence.
\newblock {\em Phys. Rev. Lett.}, 39:691---694, 1977.

\bibitem{Short1983}
R.~Short and L.~Mandel.
\newblock Observation of {sub-Poissonian} photon statistics.
\newblock {\em Phys. Rev. Lett.}, 51:384---387, 1983.

\bibitem{Teich1985}
M.~C. Teich and B.~E.~A. Saleh.
\newblock Observation of {sub-Poisson} {Franck-Hertz} light at 253.7 nm.
\newblock {\em J. Opt. Soc. Am. B}, 2:275---282, 1985.

\bibitem{Tapster1987}
P.~R. Tapster, J.~G. Rarity, and J.~S. Satchell.
\newblock Generation of {sub-Poissonian} light by high-efficiency
  light-emitting diodes.
\newblock {\em Europhys. Lett.}, 4:293---299, 1987.

\bibitem{Li1994}
R.-D. Li and P.~Kumar.
\newblock Quantum-noise reduction in travelling-wave second-harmonic
  generation.
\newblock {\em Phys. Rev. A}, 49:2157---2166, 1994.

\bibitem{Bajer1999}
J.~Bajer, O.~Haderka, and J.~Pe\v{r}ina.
\newblock {Sub-Poissonian} behaviour in the second harmonic generation.
\newblock {\em J. Opt. B: Quantum Semiclass. Opt}, 1:529---533, 1999.

\bibitem{Li1995}
R.-D. Li, S.-K. Choi, C.~Kim, and P.~Kumar.
\newblock Generation of {sub-Poissonian} pulses of light.
\newblock {\em Phys. Rev. A}, 51:R3429---R3432, 1995.

\bibitem{Koashi1993}
M.~Koashi, K.~Kono, T.~Hirano, and M.~Matsuoka.
\newblock Photon antibunching in pulsed squeezed light generated via parametric
  amplification.
\newblock {\em Phys. Rev. Lett.}, 71:1164---1167, 1993.

\bibitem{Mertz1990}
J.~Mertz, A.~Heidmann, C.~Fabre, E.~Giacobino, and S.~Reynaud.
\newblock Observation of high-intensity {sub-Poissonian} light using an optical
  parametric oscillator.
\newblock {\em Phys. Rev. Lett.}, 64:2897---2900, 1990.

\bibitem{Kim1992}
C.~Kim and P.~Kumar.
\newblock Tunable {sub-Poissonian} light generation from a parametric amplifier
  using an intensity feedforward scheme.
\newblock {\em Phys. Rev. A}, 45:5237---5242, 1992.

\bibitem{Raimond2001}
J.~M. Raimond, M.~Brune, and S.~Haroche.
\newblock Manipulating quantum entanglement with atoms and photons in a cavity.
\newblock {\em Rev. Mod. Phys.}, 73:565---583, 2001.

\bibitem{Rarity1987}
J.G. Rarity, P.R. Tapster, and E.~Jakeman.
\newblock Observation of sub-Poissonian light in parametric downconversion.
\newblock {\em Opt. Commun.}, 62:201---206, 1987.

\bibitem{Laurat2003}
J.~Laurat, T.~Coudreau, N.~Treps, A.~Maitre, and C.~Fabre.
\newblock Conditional preparation of a quantum state in the continuous variable
  regime: Generation of a {sub-Poissonian} state from twin beams.
\newblock {\em Phys. Rev. Lett.}, 91:213601, 2003.

\bibitem{Zou2006}
H.~Zou, S.~Zhai, J.~Guo, R.~Yang, and J.~Gao.
\newblock Preparation and measurement of tunable highpower {sub-Poissonian}
  light using twin beams.
\newblock {\em Opt. Lett.}, 31:1735---1737, 2006.

\bibitem{Bondani2007}
M.~Bondani, A.~Allevi, G.~Zambra, M.~G.~A. Paris, and A.~Andreoni.
\newblock Sub-shot-noise photon-number correlation in a mesoscopic twin beam of
  light.
\newblock {\em Phys. Rev. A}, 76:013833, 2007.

\bibitem{PerinaJr2013b}
J.~{Pe\v{r}ina~Jr.}, O.~Haderka, and V.~Mich\'{a}lek.
\newblock Sub-Poissonian-light generation by postselection from twin beams.
\newblock {\em Opt. Express}, 21:19387---19394, 2013.

\bibitem{Lamperti2014}
M.~Lamperti, A.~Allevi, M.~Bondani, R.~Machulka, V.~Mich\' alek,
O.~Haderka,
  and J.~{Pe\v{r}ina~Jr.}
\newblock Optimal sub-Poissonian light generation from twin beams by
  photon-number resolving detectors.
\newblock {\em JOSA B}, 31:20--25, 2014.

\bibitem{Iskhakov2016}
T.~S. Iskhakov, V.~C. Usenko, U.~L. Andersen, R.~Filip, M.~V.
Chekhova, and
  G.~Leuchs.
\newblock Heralded source of bright multi-mode mesoscopic sub-Poissonian light.
\newblock {\em Opt. Lett.}, 41:2149---2152, 2016.

\bibitem{Harder2016}
G.~Harder, T.~J. Bartley, A.~E. Lita, S.~W. Nam, T.~Gerrits, and
C.~Silberhorn.
\newblock Single-mode parametric-down-conversion states with 50 photons as a
  source for mesoscopic quantum optics.
\newblock {\em Phys. Rev. Lett.}, 116:143601, 2016.

\bibitem{Perina1991}
J.~Pe\v{r}ina.
\newblock {\em Quantum Statistics of Linear and Nonlinear Optical Phenomena}.
\newblock Kluwer, Dordrecht, 1991.

\bibitem{Saleh1978}
B.~E.~A. Saleh.
\newblock {\em Photoelectron Statistics}.
\newblock Springer-Verlag, New York, 1978.

\bibitem{Vogel2006}
W.~Vogel and D.~G. Welsch.
\newblock {\em Quantum Optics, 3rd ed.}
\newblock Wiley-VCH, Weinheim, 2006.

\bibitem{Glauber1963}
R.~J. Glauber.
\newblock Coherent and incoherent states of the radiation field.
\newblock {\em Phys. Rev.}, 131:2766---2788, 1963.

\bibitem{Sudarshan1963}
E.~C.~G. Sudarshan.
\newblock Equivalence of semiclassical and quantum mechanical descriptions of
  statistical light beams.
\newblock {\em Phys. Rev. Lett.}, 10:277---179, 1963.

\bibitem{Lee1990a}
C.~T. Lee.
\newblock Higher-order criteria for nonclassical effects in photon statistics.
\newblock {\em Phys. Rev. A}, 41:1721---1723, 1990.

\bibitem{Mista1977}
L.~Mi\v{s}ta, {V.~Pe\v{r}inov\' a}, J.~Pe\v{r}ina, and
{Z.~Braunerov\' a}.
\newblock Quantum statistical properties of degenerate parametric amplification process.
\newblock {\em Acta Phys. Pol.}, A51:739---751, 1977.

\bibitem{Hong1985b}
C.~K. Hong and L.~Mandel.
\newblock Higher-order squeezing of a quantum field.
\newblock {\em Phys. Rev. Lett.}, 54:323---325, 1985.

\bibitem{Prakash2006}
H.~Prakash and D.~K. Mishra.
\newblock Higher order {sub-Poissonian} photon statistics and their use in
  detection of {Hong} and {Mandel} squeezing and amplitude-squared squeezing.
\newblock {\em J. Phys. B: At. Mol. Opt. Phys.}, 39:2291–--2297, 2006.

\bibitem{Verma2010}
A.~Verma and A.~Pathak.
\newblock Generalized structure of higher order nonclassicality.
\newblock {\em Phys. Lett. A}, 374:1009---1020, 2010.

\bibitem{Hong1985a}
C.~K. Hong and L.~Mandel.
\newblock Generation of higher-order squeezing of quantum electromagnetic
  fields.
\newblock {\em Phys. Rev. A}, 32:974---982, 1985.

\bibitem{Hillery1987a}
M.~Hillery.
\newblock Amplitude-squared squeezing of the electromagnetic field.
\newblock {\em Phys. Rev. A}, 36:3796---3802, 1987.

\bibitem{Hillery1987b}
M.~Hillery.
\newblock Squeezing of the square of the field amplitude in second harmonic
  generation.
\newblock {\em Opt. Commun.}, 62:135---138, 1987.

\bibitem{Kim1998}
K.~Kim.
\newblock Higher order {sub-Poissonian}.
\newblock {\em Phys. Lett. A}, 245:40---42, 1998.

\bibitem{Klyshko1996}
D.~N. Klyshko.
\newblock Observable signs of nonclassical light.
\newblock {\em Phys. Lett. A}, 213:7---15, 1996.

\bibitem{Lee1998}
C.~T. Lee.
\newblock Simple criterion for nonclassical two-mode states.
\newblock {\em J. Opt. Soc. Am. B}, 15:1187---1191, 1998.

\bibitem{Arkhipov2016c}
I.~I. Arkhipov, J.~{Pe\v{r}ina~Jr.}, V.~Mich\'{a}lek, and
O.~Haderka.
\newblock Experimental detection of nonclassicality of single-mode fields via
  intensity moments.
\newblock {\em Opt. Express}, 24:29496---29505, 2016.

\bibitem{Lee1991}
C.~T. Lee.
\newblock Measure of the nonclassicality of nonclassical states.
\newblock {\em Phys. Rev. A}, 44:R2775---R2778, 1991.

\bibitem{Gradshtein2000}
I.~S. Gradshtein and I.~M. Ryzhik.
\newblock {\em Table of Integrals, Series, and Products, 6th ed.}
\newblock Academic Press, San Diego, 2000.

\bibitem{PerinaJr2012}
J.~{Pe\v{r}ina~Jr.}, M.~Hamar, V.~Mich\'{a}lek, and O.~Haderka.
\newblock Photon-number distributions of twin beams generated in spontaneous
  parametric down-conversion and measured by an intensified {CCD} camera.
\newblock {\em Phys. Rev. A}, 85:023816, 2012.

\bibitem{PerinaJr2013a}
J.~{Pe\v{r}ina~Jr.}, O.~Haderka, V.~Mich\'{a}lek, and M.~Hamar.
\newblock State reconstruction of a multimode twin beam using photodetection.
\newblock {\em Phys. Rev. A}, 87:022108, 2013.

\bibitem{PerinaJr2012a}
J.~{Pe\v{r}ina~Jr.}, O.~Haderka, M.~Hamar, and V.~Mich\'{a}lek.
\newblock Absolute detector calibration using twin beams.
\newblock {\em Opt. Lett.}, 37:2475---2477, 2012.

\bibitem{Perina2005}
J.~Pe\v{r}ina and J.~K\v{r}epelka.
\newblock Multimode description of spontaneous parametric down-conversion.
\newblock {\em J. Opt. B: Quant. Semiclass. Opt.}, 7:246---252, 2005.

\bibitem{Dempster1977}
A.~P. Dempster, N.~M. Laird, and D.~B. Rubin.
\newblock Maximum likelihood from incomplete data via the {EM} algorithm.
\newblock {\em J. R. Statist. Soc. B}, 39:1---38, 1977.

\end{thebibliography}
% \bibliographystyle{unsrt}

\end{document}